\documentclass[sigconf]{acmart}

\AtBeginDocument{%
  }

\setcopyright{acmlicensed}
\copyrightyear{2025}
\acmYear{2025}
\setcopyright{acmlicensed}\acmConference[WWW Companion '25]{Companion
Proceedings of the ACM Web Conference 2025}{April 28-May 2, 2025}{Sydney,
NSW, Australia}
\acmBooktitle{Companion Proceedings of the ACM Web Conference 2025 (WWW
Companion '25), April 28-May 2, 2025, Sydney, NSW, Australia}
\acmDOI{10.1145/3701716.3715227}
\acmISBN{979-8-4007-1331-6/2025/04}

\usepackage{mathtools}
\usepackage{amsfonts}
\usepackage{bm}
\usepackage{amsmath}
\usepackage{amssymb}
\usepackage{booktabs}
\usepackage{tabularx}
\usepackage{algorithm}
\usepackage{algorithmic}
\newcommand{\paragraphseciton}[1]{\vspace{0.5em}\noindent\textbf{#1}\hspace{0.1em}}
\DeclarePairedDelimiter{\norm}{\lVert}{\rVert}

\usepackage{graphicx} 
\usepackage{multirow} 
\usepackage{pifont}
\usepackage{stfloats}
\begin{document}

\title{Generalized Contrastive Learning for Multi-Modal \\ Retrieval and Ranking}

\author{Tianyu Zhu}
\affiliation{%
  \institution{marqo.ai}
  \city{Melbourne}
  \state{Victoria}
  \country{Australia}
}
\email{tianyuzhu52@gmail.com}
\authornote{Corresponding author}

\author{Myong Chol Jung}
\affiliation{%
  \institution{marqo.ai}
  \city{Melbourne}
  \state{Victoria}
  \country{Australia}
}
\email{david@marqo.ai}

\author{Jesse Clark}
\affiliation{%
  \institution{marqo.ai}
  \city{Melbourne}
  \state{Victoria}
  \country{Australia}
}
\email{jesse@marqo.ai}
\renewcommand{\shortauthors}{Tianyu Zhu, Myong Chol Jung, and Jesse Clark}

\begin{abstract}
Contrastive learning has gained widespread adoption for retrieval tasks due to its minimal requirement for manual annotations. 
However, popular training frameworks typically learn from binary (positive/negative) relevance, making them ineffective at incorporating desired rankings. As a result, the poor ranking performance of these models forces systems to employ a re-ranker, which increases complexity, maintenance effort and inference time. 
To address this, we introduce Generalized Contrastive Learning (GCL), a training framework designed to learn from continuous ranking scores beyond binary relevance. 
GCL encodes both relevance and ranking information into a unified embedding space by applying ranking scores to the loss function. This enables a single-stage retrieval system.
In addition, during our research, we identified a lack of public multi-modal datasets that benchmark both retrieval and ranking capabilities. To facilitate this and future research for ranked retrieval, we curated a large-scale MarqoGS-10M dataset using GPT-4 and Google Shopping, providing ranking scores for each of the 10 million query-document pairs.
Our results show that GCL achieves a \textbf{29.3\%} increase in NDCG@10 for in-domain evaluations and \textbf{6.0\% to 10.0\%} increases for cold-start evaluations compared to the finetuned CLIP baseline with MarqoGS-10M. Additionally, we evaluated GCL offline on a proprietary user interaction data. GCL shows an \textbf{11.2\%} gain for in-domain evaluations. 
The dataset and the method are available at: \href{https://github.com/marqo-ai/GCL}{https://github.com/marqo-ai/GCL}.

\end{abstract}



\maketitle

\section{Introduction}
\begin{figure*}[t]
  \centering
  \includegraphics[width=0.95\linewidth]{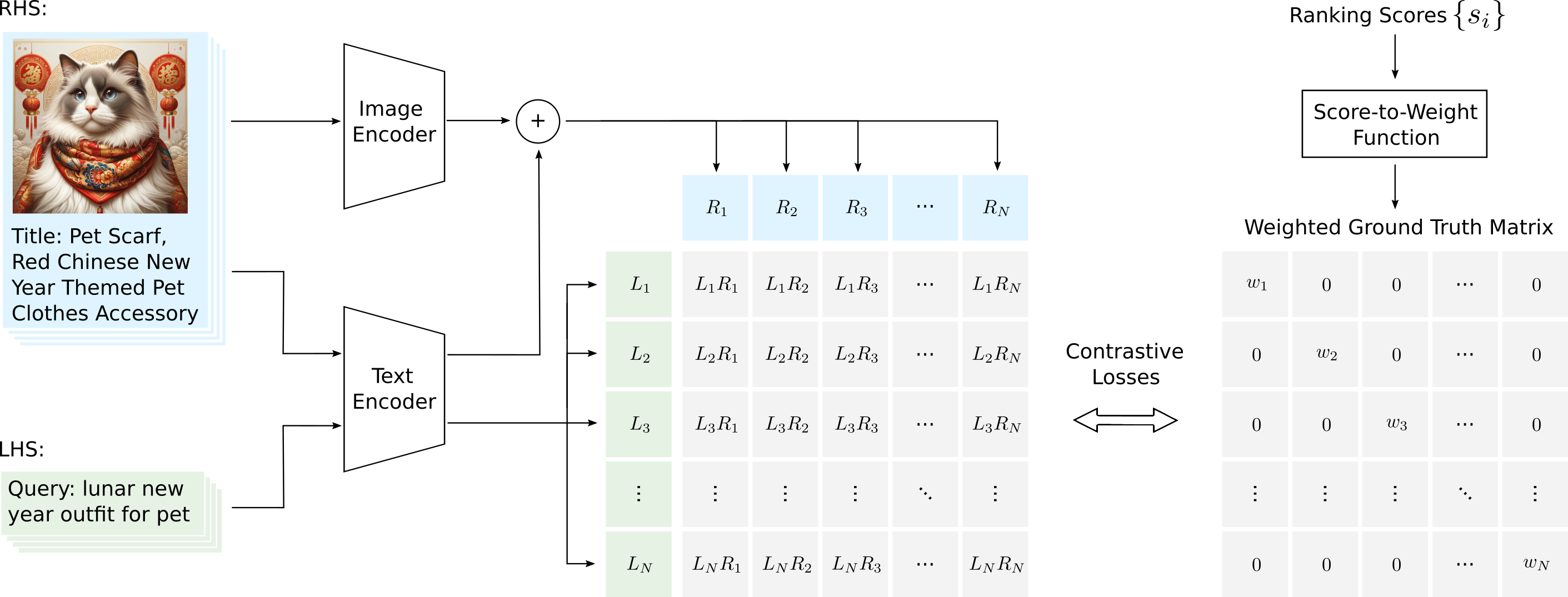}
  \caption{Overview of the Generalized Contrastive Learning. GCL integrates ranking information alongside multiple input fields for each sample across both left-hand-side (LHS) and right-hand-side (RHS). Ground-truth ranking scores are transformed into weights, which are used for computing contrastive losses, ensuring that pairs with higher weights incur greater penalties.}
  \label{fig: main}
\end{figure*}

Recently, latent representations learned via contrastive learning have gained significant adoption for cross-modal retrieval tasks within the research and vector database community~\cite{yuan2021florence,ma2022ei,yu2022coca,huaishao2022clip4clip, liu2019roberta}. These methods have effectively replaced earlier retrieval techniques such as BM25~\cite{robertson1995okapi}.
However, despite their popularity, existing contrastive methods~\cite{chen2020simple, schroff2015facenet, sohn2016improved, gutmann2010noise, oord2018representation} were initially designed for pretraining foundation models and have significant shortcomings when directly applied to fine-tuning on a retrieval dataset. 
They assume a one-to-one mapping between queries and documents, and learn from binary (positive/negative) relevance~\cite{radford2021learning, jia2021scaling, jaiswal2021survey}. Therefore, these frameworks cannot train the model to rank relevant documents if explicit desired ranking orders are available. As a result, a second-stage re-ranker becomes necessary for these models~\cite{nogueira2020document,asadi2013effectiveness}, increasing system complexity and inference time. 

To address this shortcoming, we propose \textbf{G}eneralized \textbf{C}ontrastive \textbf{L}earning for Multi-Modal Retrieval and Ranking (GCL), a training framework that integrates detailed relevance and ranking information. Unlike traditional contrastive learning approaches~\cite{radford2021learning,jia2021scaling,oord2018representation} that rely on pairs of queries and documents with binary relevance, our framework generalizes this by incorporating a continuous weight for each pair, thus creating a triplet input unit. The weights are converted from ground-truth ranking score by a score-to-weight function. During training, the weights scale the loss associated with each pair, ensuring that embeddings of documents with higher ranking scores are pushed closer to the queries by the gradients. 
Fine-tuning models using historical user interaction data is common in industrial retrieval systems. GCL enables the model to learn ranking information directly from such data. For example, if many users have purchased a product after searching for a particular query, we can assign a higher weight to that query–document pair. Consequently, the model learns to produce higher similarity scores for this pair, promoting it to the top of the search results.
Moreover, we extend traditional single-field learning by training with multiple fields in GCL, merging elements such as the title and product image into a weighted average embedding. By incorporating information from both text and visual modalities, we create more comprehensive document representations that enhance retrieval performance.

During our research, we discovered a significant lack of public datasets that include ranking information essential for benchmarking retrieval models. A detailed discussion of the public retrieval datasets is included in Section~\ref{rw: dataset}. We believe this scarcity has greatly hindered research progress in this field.
Meaningful rankings are typically derived from historical human interactions~\cite{huang2013learning, gao2010clickthrough, gao2011clickthrough}, such as click-through data from Google, add-to-cart events on Amazon, or engagement rates on YouTube. However, manual annotation of such data is infeasible at scales, and user interaction logs are proprietary. Consequently, the absence of public datasets with ranking information poses a significant barrier to benchmarking ranked retrieval models.

Since the underlying search logs are unavailable, we posit that the listing positions of products returned by services like Google Shopping are derived from these logs and provide meaningful ranked mappings between queries and documents. Therefore, we propose using these listing positions as proxies for ranking scores to benchmark the capability of training frameworks to learn from them. In this paper, we curated MarqoGS-10M: a large-scale multimodal dataset comprising 10 million query–document pairs, each accompanied by a ranking score from 1 to 100 derived from their listing positions. We generated approximately 100,000 unique queries using GPT-4~\cite{achiam2023gpt} in a hierarchical manner based on Amazon categories to ensure quality. Then, we retrieved the top 100 products per query from Google Shopping, with each product accompanied by an image and a title text. This dataset has been partitioned into in-domain, novel queries, novel documents and zero-shot sets, providing precise insights for benchmarking the models. The latter three subsets are collectively referred to as cold-start evaluations.

Compared to the seminal contrastive method CLIP~\cite{radford2021learning}, ViT-L/14 trained with GCL shows a \textbf{29.3\%} increase in NDCG@10 and a \textbf{46.9\%} increase in ERR for in-domain evaluation. For cold-start evaluations, it exhibits relative improvements of \textbf{6.0 - 10.0\%} in NDCG@10, \textbf{3.5 - 8.1\%} in ERR, and \textbf{5.7 - 8.6\%} in RBP. The large gap between GCL and the CLIP baseline indicates significant potential for future research. The improvement in non-training set evaluations validates that the ranking in the dataset meaningfully captures the underlying mapping between queries and documents. In addition to MarqoGS-10m, we perform an offline evaluation of GCL with our proprietary user interaction data collecting from an ecommerce platform. GCL showed an \textbf{11.2\%} gain for in-domain evaluations. 

In summary, our contributions are as follows:
\begin{itemize}
    \item We propose a novel contrastive learning framework that generalizes beyond binary relevance learning to accommodate fine-grained rankings.
    \item We expand the conventional single-field representation of queries and documents to encompass multiple fields.
    \item We compile a large-scale multi-modal ranked retrieval dataset with ranking scores
    \item We introduce an innovative split of the dataset that facilitates comprehensive evaluation insights.
\end{itemize}

\section{Related Works}
\renewcommand{\arraystretch}{1.35}
\setlength{\tabcolsep}{5pt}
\begin{table*}[t]

\centering
\small
\caption{Statistical overview of popular datasets used in retrieval tasks, detailing rankings, dataset modality, average documents per query, total number of queries, and corpus size. This demonstrates the unique contribution of the Marqo-GS-10M dataset, featuring multimodality with ranking scores, a high document-to-query ratio across large scale queries and corpus.}
\begin{tabular}{l|c|c|c|c|c|c}
\toprule
\textbf{Dataset} & \textbf{Rankings} & \textbf{Multimodal} & \textbf{Dataset Split} & \textbf{Avg. D/Q} & \textbf{Total \#Corpus} & \textbf{Total \#Query} \\
\hline
MSMarco~\cite{nguyen2016ms} & \ding{55} & \ding{55} & Train/Test & 1.1 & 8.84m & 1.01m \\

NQ~\cite{kwiatkowski2019natural} & \ding{55} & \ding{55} & Train/Test & 1.2 & 2.68m & 323k\\

Trec-Covid~\cite{voorhees2021trec} & 3-level & \ding{55} & Test & 493.5 & 171k & 50 \\

NFCorpus~\cite{boteva2016full} & 3-level & \ding{55} & Train/Dev/Test & 38.2 & 3.63k & 3.24k \\



TREC-NEWS~\cite{craswell2020overview} & 5-level & \ding{55} & Test & 19.6 & 595k & 57 \\

Robust04~\cite{voorhees2003overview} & 3-level & \ding{55} & Test & 69.9 & 528k & 249\\



FEVER~\cite{thorne2018fever} & \ding{55} & \ding{55} & Train/Dev/Test & 1.2 & 5.42m & 130k\\

SciFact~\cite{wadden2020fact} & \ding{55} & \ding{55} & Train/Test & 1.1 & 5.18k & 1.4k\\


Signal-1M~\cite{suarez2018data} & 3-level & \ding{55} & Test & 19.6 & 2.87m & 97\\


CQADupStack~\cite{ioffe2010improved} & \ding{55} & \ding{55} & Test & 1.4 & 457k & 13.1k\\

Touche-2020~\cite{bondarenko2020overview} & 3-level & \ding{55} & Test & 19.0 & 383k & 49\\

Climate-FEVER~\cite{diggelmann2020climate} & \ding{55} & \ding{55} & Test & 3.0 & 5.42m & 1.54k\\

\hline
ImageNet~\cite{deng2009imagenet} & \ding{55} & \ding{51} & Train/Dev/Test & 1.43k & 1.43m & 1k  \\

COCO~\cite{lin2014microsoft} & \ding{55} & \ding{51} & Train/Dev/Test & \textasciitilde1.0 & 330k & 1.5m \\

Flickr30k~\cite{plummer2015flickr30k} & \ding{55} & \ding{51} & Train/Dev/Test & \textasciitilde1.0 & 31.8k & 158k \\

LAION-400M~\cite{schuhmann2021laion} & \ding{55} & \ding{51} & - & \textasciitilde1.0 & \textasciitilde400m & \textasciitilde400m \\

Visual Genome~\cite{krishna2017visual} & \ding{55} & \ding{51} & Train/Dev/Test & \textasciitilde1.0 & 108k & \textasciitilde5.40m \\

RedCaps~\cite{desai2021redcaps} & \ding{55} & \ding{51} & - & \textasciitilde1.0 & \textasciitilde12m & \textasciitilde12m \\

CC12M~\cite{changpinyo2021conceptual} & \ding{55} & \ding{51} & - & \textasciitilde1.0 & \textasciitilde12m & \textasciitilde12m \\

\hline
MarqoGS-10M (Ours) & \ding{51} & \ding{51} & Quadruple-Split + Dev/Test & 100 & 5.50m & 98.2k\\
\bottomrule
\end{tabular}
\label{tab: dataset}
\end{table*}

\subsection{Contrastive learning}
Generative learning ~\cite{radford2018improving, le2013building, goodfellow2020generative} and contrastive learning~\cite{chen2020simple, schroff2015facenet, radford2021learning} are the prevailing methods to learn effective visual and text embeddings without manual annotation. Among them, contrastive learning benefits from spatial proximity of semantically similar objects, making it a widely adopted paradigm for learning representation in text and cross-modal retrieval tasks~\cite{wang2022text, ray2024cola, radford2021learning}. The spatial proximity of similar objects is achieved by minimizing their distance in the vector space while introducing negative samples to avoid mode collapsing~\cite{sohn2016improved, gutmann2010noise, oord2018representation, schroff2015facenet, chen2020simple, kalantidis2020hard}. 
The use of positive/negative pairs has proven effective for learning text embeddings~\cite{wang2022text, reimers2019sentence, liu2019roberta}, image embeddings~\cite{chen2020simple, tian2020contrastive, he2020momentum}, and cross-modal embeddings~\cite{radford2021learning, elizalde2023clap, yuan2021florence, ma2022ei, yu2022coca, huaishao2022clip4clip, zhai2023sigmoid}. Models pretrained contrastively have also shown improvement for downstream tasks such as object detection~\cite{yuan2021florence,vidit2023clip}, image captioning~\cite{yu2022coca,mokady2021clipcap}, and few-shot learning~\cite{zhou2022learning,zhou2022conditional}. Despite being the default learning paradigm for retrieval, the current contrastive learning methods are limited in their capacity to explicitly learn the rank order of documents given a query, thus constraining their utility in rank optimization. Additionally, these methods are  constrained by their focus on maximizing similarity between individual instances (i.e., one-to-one similarity), thereby limiting the exploration of similarity relationships between sets of instances (i.e., many-to-many similarity).

\subsection{Information Retrieval Datasets}
\label{rw: dataset}
Information Retrieval (IR) datasets are designed to assess the retrieval and ranking capabilities of various models. However, comprehensive ranking information is notably lacking in most available datasets. Popular text datasets such as MSMARCO~\cite{nguyen2016ms}, NQ~\cite{kwiatkowski2019natural} and FEVER~\cite{thorne2018fever} contain large-scale corpus with sufficient number of unique queries, but they only capture a binary relationship between the queries and the documents. Some text datasets such as TREC-Covid~\cite{voorhees2021trec}, TREC-News~\cite{craswell2020overview} and NFCorpus~\cite{boteva2016full} offer non-binary relevance but they are limited to three to five relevance level while having small number of unique queries as shown in Table~\ref{tab: dataset}. The matter got worse for cross-modal text to image retrieval, where popular methods use image captioning datasets such as COCO~\cite{lin2014microsoft}, Flickr30k~\cite{plummer2015flickr30k} and Visual Genome~\cite{krishna2017visual} to benchmark their retrieval performances. These image caption datasets along with other large-scale cross-modal datasets such as LAION-400M~\cite{schuhmann2021laion}, RedCaps~\cite{desai2021redcaps} and CC12M~\cite{changpinyo2021conceptual} typically assumes a one-to-one mapping between text and image without rankings. They are designed for cross-modal pretraining and captioning but not representative of retrieval tasks in general. 
Another limitation of these datasets is their division into train, dev and test~\cite{thakur2021beir}. This conventional split does not reflect real-world challenge. It lacks detailed evaluations on novel queries with existing documents, novel documents with existing queries, or completely zero-shot query-document pairs. 

\subsection{Neural information retrieval}
Neural information retrieval aims to locate and retrieve relevant documents corresponding to queries by capturing semantic relationships between the queries and the documents learned from deep learning models~\cite{zhao2022dense}. 
Recent advancements of neural IR have been made in sparse retrieval methods to effectively reduce data dimensionality~\cite{jang2021ultra,lassance2021composite}, thereby enhancing retrieval efficiency via various methods such as term re-weighting~\cite{frej2020learning,mackenzie2020efficiency} and expansion methods~\cite{nogueira2019document,yan2021unified,bonifacio2022inpars,jeronymo2023inpars}. Similarly, significant advancements  have been made in the domain of dense retrieval~\cite{yu2021improving,khattab2020colbert,santhanam2021colbertv2} by utilizing pre-trained language models like BERT~\cite{kenton2019bert,nie2020dc}. 
Despite the notable advancements achieved by these studies, the majority require a reranking stage after the initial retrieval. Our work takes a different approach by learning the retrieval and ranking simultaneously, introducing a unified single-stage retrieval system that optimizes business metrics more directly than two-stage systems. 

\section{Generalized Contrastive Learning}
\subsection{Incorporating Ranking Signals to Contrastive Learning}
\label{sec: weighted-clip}

In this section, we present generalized contrastive learning (GCL), a novel framework that integrates ranking signals into the contrastive learning process by utilizing weights. Traditional contrastive learning techniques rely on a dataset comprising $N_{D}$ pairs of ${(q_i, d_i)}$, where $q_i$ and $d_i$ denote the query and document for the $i^{th}$ sample. 

\begin{figure}[t]
\centering
\includegraphics[width=0.9\linewidth]{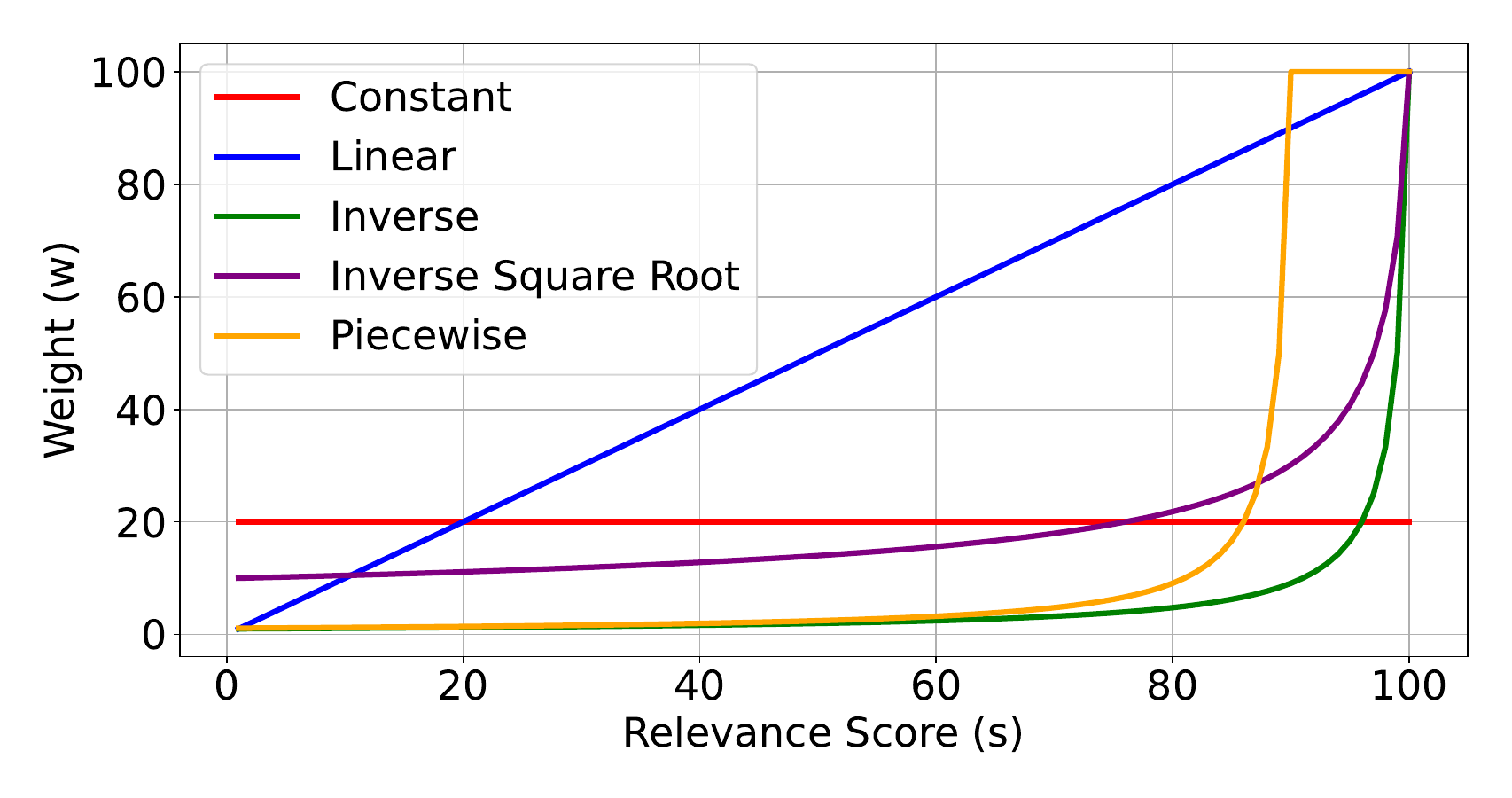}
\caption{Plot of various score-to-weight functions. }

\label{fig: stw}
~\vspace{-2em}
\end{figure}

In the context of the original CLIP framework~\cite{radford2021learning}, texts can be considered as queries and images as corresponding documents. 
In contrast, our method employs a dataset consisting of $N_{D}$ triplets of $(q_i, d_i, w_i)$, where $w_i$ represents the weight. These weights are derived from desired relevance or ranking scores $s_i$ by a Score-to-Weight function $w_i = STW(s_i)\in \mathbb{R}$. STW functions are used to shape the distribution of weights to help improving certain metrics. The higher score $s_i$ of a document $d_i$, the greater its corresponding weight $w_i$ will be. In our study, we have experimented with five different STW functions namely: constant, linear, inverse, inverse square root, and piecewise functions. Given a score $s$ of an instance, the maximum of possible score $s_{max}$, and a constant $c$, the functions are defined as: 

\begin{align}
    \label{eq: STW}
    &\textbf{Constant: } c, \quad\textbf{Linear: } s, \quad\textbf{Inverse: } \frac{s_{max}}{s_{max}-s+1} \\
    &\textbf{Inverse square root: } \frac{s_{max}}{\sqrt{s_{max}-s+1}} \\
    &\textbf{Piecewise: } \begin{cases} s_{max} & s \geq 0.9 s_{max}, \\ \frac{s_{max}}{0.9 s_{max} - s + 1} & s < 0.9 s_{max}  \end{cases}.
\end{align}

When the constant function is used with $c=1$, the weighted contrastive losses return to the original CLIP loss. The plot of the STW functions can be seen in Figure~\ref{fig: stw}.
For each training iteration, a batch of $N$ triplets $(\mathbf{Q}, \mathbf{D}, \mathbf{w})=(\{q_i\}_{i=1}^N, \{d_i\}_{i=1}^N, \{w_i\}_{i=1}^N)$ is randomly sampled. 
Query encoder $E_q$ and document encoder $E_d$ then encode the queries and documents into $k\text{-dimensional}$ embeddings $\mathbf{Q_f}\in\mathbb{R}^{N\times k}$ and $\mathbf{D_f}\in\mathbb{R}^{N\times k}$ respectively. 
The embeddings are then normalized as $\mathbf{\hat{Q}_f}=\mathbf{Q_f}/\norm{\mathbf{Q_f}}_2$ and 
$\mathbf{\hat{D}_f}=\mathbf{D_f}/\norm{\mathbf{D_f}}_2$ where $\norm{\cdot}_2$ is the $l^2\text{-norm}$.
The encoders can be text or image encoders depending on the data type of the queries and documents. The pairwise dot product results in $\mathbf{Z} = \mathbf{\hat{Q}_f} \cdot \mathbf{\hat{D}_f}^T \in \mathbb{R}^{N\times N}$, capturing the similarity scores between each query-document pair within the batch. The loss is computed from $\mathbf{Z}$ and $\mathbf{w}$ using weighted cross-entropy loss.

\subsection{Weighted Contrastive Loss}
\label{subsec: weighted contrastive losses}
Once the pairwise dot product matrix $\mathbf{Z}$ is calculated, we apply a loss function designed to penalize low values on the diagonal and high values off the diagonal. The diagonal entries of $\mathbf{Z}$ reflect the dot products of matching query-document pairs, indicating relevance. Off-diagonal entries, which represent dot products of non-matching pairs within the batch, serve as in-batch negatives. For CLIP and similar contrastive learning methods~\cite{radford2021learning,jia2021scaling,yu2022coca}, the ground truth matrix is effectively an identity matrix, treating all diagonal values equally and neglecting the varying degrees of relevance between queries and documents. In our approach, we calculate this loss considering the weights $\mathbf{w}$, as depicted in Figure~\ref{fig: main}, to account for the relevance differences. 
%
We can treat the contrastive learning task as an $n$-class, $n$-sample classification problem, computing cross-entropy loss between the dot product matrix $\mathbf{Z}$ and an identity matrix. Here, the $i^{th}$ class corresponding to the $i^{th}$ sample is considered the ground truth. To infuse ranking information, our approach utilizes weighted cross-entropy loss, 
\begin{align}
 \scriptsize
  \mathcal{L}_{WCE} \left(\mathbf{Z}, \mathbf{w}\right) = - \frac{1}{2N}\Bigg(\sum_{i=1}^{N} w_i \log&\left(\frac{\exp(\mathbf{Z}[i,i])}{\sum_{j=1}^N \exp(\mathbf{Z}[i,j])}\right) + \nonumber \\ 
  &\sum_{i=1}^{N} w_i \log\left(\frac{\exp(\mathbf{Z}[i,i])}{\sum_{j=1}^N \exp(\mathbf{Z}[j,i])}\right)\Bigg)
  \label{eq: weighted ce}
\end{align}
where $\mathbf{Z}[i,j]$ denotes the element in the $i^{th}$ row and the $j^{th}$ column of $\mathbf{Z}$. This variation of cross-entropy loss assigns greater penalties to rows and columns with higher weights, biasing the gradients towards prioritizing their correction. This approach ensures that pairs deemed more relevant based on their weights are adjusted preferentially during the training process.
\begin{algorithm}[t]
   \caption{Single-Field GCL}

   \label{alg: wclip}
\begin{algorithmic}[1]
   \STATE {\bfseries Input:} A batch of $N$ triplets $(\mathbf{Q}, \mathbf{D}, \mathbf{w})$ of queries, documents and weights.

   \STATE Compute $\mathbf{Q_f} = E_q(\mathbf{Q})$ and $\mathbf{D_f} = E_d(\mathbf{D})$ with a query encoder $E_q$ and a document encoder $E_d$.

    \STATE Normalize $\mathbf{Q_f}$ and $\mathbf{D_f}$ to unit vectors $\mathbf{\hat{Q}_f}$ and $\mathbf{\hat{D}_f}$.
    \STATE Compute dot product $\mathbf{Z} = \mathbf{\hat{Q}_f} \cdot \mathbf{\hat{D}_f}^T$.

    \STATE Compute loss $\mathcal{L}=\mathcal{L}_{WCE}(\mathbf{Z}, \mathbf{w})$ in Eq.~\eqref{eq: weighted ce}.

    \STATE Back propagate $\mathcal{L}$ to update $E_q$ and $E_d$.
    
\end{algorithmic}
\end{algorithm}

\begin{algorithm}[tb]
   \scriptsize
   \caption{Multi-Field GCL}
   \label{alg: multi-field}
\begin{algorithmic}[1]
   \STATE {\bfseries Input:} A batch of $N$ triplets $(\mathbf{L}, \mathbf{R}, \mathbf{w})$, which are LHS fields, RHS fields, and weights. Hyperparameters $\bm{\gamma}_L$ and $\bm{\gamma}_R$ representing weights of each field. The number of fields in LHS as $m$ and the number of fields in RHS as $n$.
   \STATE Compute $\mathbf{L}^f=\{E_j(\mathbf{L}^j)\}_{j=1}^m$ and $\mathbf{R}^f=\{E_j(\mathbf{R}^j)\}_{j=1}^n$ with a field-specific encoder $E_j$.
  \STATE Normalize embeddings of each field in $\mathbf{L}^f$ and $\mathbf{R}^f$ to obtain $\mathbf{\hat{L}}^f$ and $\mathbf{\hat{R}}^f$.
  \STATE Compute weighted average embeddings $\mathbf{\hat{L}}^f_{avg}=\sum_{j=1}^m \gamma_{Lj} \times \mathbf{\hat{L}}^f_j$ and $\mathbf{\hat{R}}^f_{avg}=\sum_{j=1}^n \gamma_{Rj} \times \mathbf{\hat{R}}^f_j$.
  \STATE Compute dot product between the averaged embeddings $\mathbf{Z}_{avg}=\mathbf{\hat{L}}^f_{avg}\cdot (\mathbf{\hat{R}}^f_{avg})^T$.
  \STATE Compute dot product between each field of LHS and RHS \\ $\{\{\mathbf{Z}^{LR}_{jk}\}_{j=1}^m\}_{k=1}^n=\{\{\mathbf{\hat{L}}^f_j\cdot (\mathbf{\hat{R}}^f_k)^T\}_{j=1}^m\}_{k=1}^n$.
  \STATE Compute loss $\mathcal{L}=\mathcal{L}_{WCE}(\mathbf{Z}_{avg},\mathbf{w})+\sum_{j=1}^m \sum_{k=1}^n \mathcal{L}_{WCE}(\mathbf{Z}^{LR}_{jk},\mathbf{w})$.
  \STATE Back propagate $\mathcal{L}$ to update all encoders.
\end{algorithmic}
\end{algorithm}

\subsection{Multi-Field}
\label{subsec: multi-field}
Previous contrastive learning approaches~\cite{radford2021learning,jia2021scaling,oord2018representation,yu2022coca,jia2021scaling} typically employ a single field to represent either a query or a document.
Our framework extends contrastive learning to multi-field, allowing both queries and documents to be represented by multiple text and image fields. This approach mirrors real-world scenarios closely, where a document often has multiple fields such as a title, image, and description. 
To distinguish from the previous single field formulation, we now denote 
a batch of $N$ triplets more generally as $(\mathbf{L}, \mathbf{R}, \mathbf{w})=\left(\{\mathbf{L}^j\}_{j=1}^m, \{\mathbf{R}^j\}_{j=1}^n, \mathbf{w}\right)$, which are left-hand-side fields (LHS), right-hand-side fields (RHS) and weights where $m$ is the number of fields in LHS,  $n$ is the number of fields in RHS, $\mathbf{L}^j$ denotes $N$ samples of the $j^{th}$ field in LHS, and $\mathbf{R}^j$ denotes $N$ samples of the $j^{th}$ field in RHS.
During training, the data from each field are processed by their respective encoders $E_j$ to extract embeddings as $\mathbf{L}^f=\{E_j(\mathbf{L}^j)\in\mathbb{R}^{N\times k}\}_{j=1}^m$ and $\mathbf{R}^f=\{E_j(\mathbf{R}^j)\in\mathbb{R}^{N\times k}\}_{j=1}^n$. Embeddings of each field are then normalized as done in Section~\ref{sec: weighted-clip} resulting in $\mathbf{\hat{L}}^f$ and $\mathbf{\hat{R}}^f$. Subsequently, a weighted average embedding is computed as $\mathbf{\hat{L}}^f_{avg}=\sum_{j=1}^m \gamma_{Lj} \times \mathbf{\hat{L}}^f_j$ and $\mathbf{\hat{R}}^f_{avg}=\sum_{j=1}^n \gamma_{Rj} \times \mathbf{\hat{R}}^f_j$ where $\bm{\gamma}_L=\{\gamma_{Lj}\}_{j=1}^m$ and $\bm{\gamma}_R=\{\gamma_{Rj}\}_{j=1}^n$ are the predetermined weights that sum to 1, and $\mathbf{\hat{L}}^f_j$ and $\mathbf{\hat{R}}^f_j$ represent the normalized embeddings of the $j^{th}$ field.
Finally, the dot product is computed by $\mathbf{Z}_{avg}=\mathbf{\hat{L}}^f_{avg}\cdot (\mathbf{\hat{R}}^f_{avg})^T \in \mathbb{R}^{N\times N}$.

\begin{figure*}[t]
\centering
\includegraphics[width=0.9\linewidth]{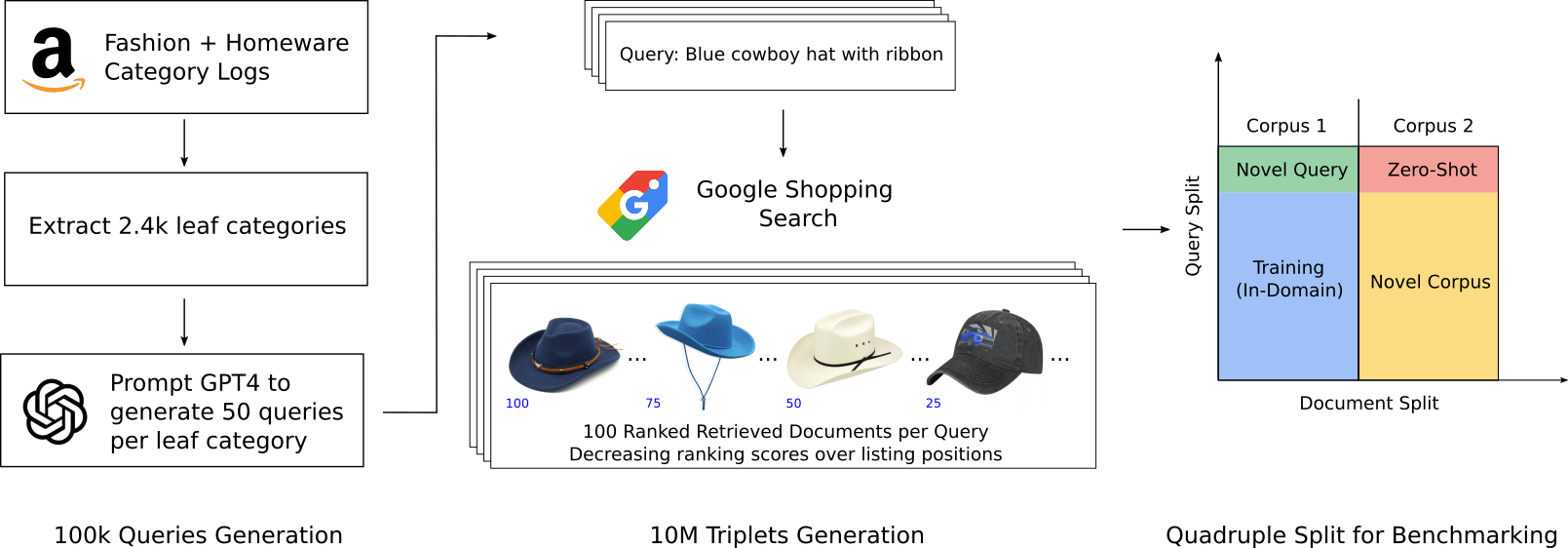}
\caption{Overview of the MarqoGS-10M data curation and Quadruple Split. We first extract 2.4k leaf categories from Amazon fashion and homeware sections, which we use to prompt GPT-4 for query generation. Each query is used to retrieve 100 relevant documents via the Google Shopping API, with their listing positions converted to ranking scores, culminating in 10 million triplets. Finally, the data is split in a quadruple way that reflects real world search system.}

\label{fig: dp}
\end{figure*}

While the weighted mean embeddings of LHS and RHS serve to semantically represent the document, relying solely on the loss computed from dot product $\mathbf{Z}_{avg}$ leads to significant performance degradation when searching with a single field query or when the document contains only text or image fields. This decline is attributed to the model being trained exclusively with mean weighted embeddings. To mitigate this issue, we compute pairwise dot products between each field on the LHS and each field on the RHS as $\{\{\mathbf{Z}^{LR}_{jk}\}_{j=1}^m\}_{k=1}^n$ where $\mathbf{Z}^{LR}_{jk}=\mathbf{\hat{L}}^f_j\cdot (\mathbf{\hat{R}}^f_k)^T \in \mathbb{R}^{N\times N}$. The overall algorithm is presented in Algorithm~\ref{alg: multi-field}. 
Subsequently, we compute loss for training the multi-field generalized-CLIP:
\begin{equation}
    \mathcal{L}=\mathcal{L}_{WCE}(\mathbf{Z}_{avg},\mathbf{w})+\sum_{j=1}^m \sum_{k=1}^n \mathcal{L}_{WCE}(\mathbf{Z}^{LR}_{jk},\mathbf{w}).
\end{equation}

\section{MarqoGS-10M dataset and benchmark}
As discussed in Section~\ref{rw: dataset}, 
Public datasets do not encapsulate the ranking complexities present in real-world search scenarios, such as those found on e-commerce platforms. To investigate and tackle this problem, it is essential to obtain a multi modal dataset that focuses on the one-to-many query document relationship and includes rankings of the relevant documents. 

In this paper, We refer to the acquired dataset as \textbf{MarqoGS-10M} which consists of \textbf{GSFashion-5M} and \textbf{GSHomeware-5M}.
In this study, we chose to collect data via Google Shopping searches, as the returned product listings provide both images and texts, accompanied by meaningful rankings.  The over data curation process is demonstrated in Figure~\ref{fig: dp}. Dataset is available for download or visualization in \href{https://huggingface.co/datasets/Marqo/marqo-GS-10M}{https://huggingface.co/datasets/Marqo/marqo-GS-10M}.

\vspace{-2em}
\subsection{Queries, Documents and Rankings}
\paragraphseciton{Queries.}
For constructing a retrieval dataset through Google Shopping, achieving broad query coverage was critical. We focused on Fashion and Homeware as our main categories and identified 2.4k leaf categories using a taxonomy derived from Amazon. We then utilized GPT-4~\cite{achiam2023gpt} to craft 50 queries for each leaf category, ensuring a variety in word lengths. This process yielded around 120k queries, from which we randomly selected around 100k (98,236) for conducting searches on Google Shopping. We selected the Fashion category as it offers an excellent case study for multimodality, where both images and texts are crucial in conveying product information. The Homeware category was also chosen to facilitate a comparison, highlighting the unique aspects and challenges of multimodal information retrieval across different domains. 

\paragraphseciton{Documents.}
For our dataset, we utilize the Google Shopping API provided by SerpAPI to search for the queries. Each search yields 100 products. Same products can be returned by search different queries resulting a many-to-many mapping. The data for each product includes the title, a thumbnail link, and the product's ranking position. We acquire the thumbnail image with wget tool. 

\paragraphseciton{Relevance/Ranking Scores.}
The ranking scores for GCL could be based on historic search logs where query-document pairs can be rated by their add-to-cart, click-through, or engagement rates. However, these metrics are often confidential and unavailable for public research. Consequently, we use the product listing position from Google Shopping searches as a proxy. 
To calculate the scores, we compute $s = 101 - rank$, so that it ranges from 1 to 100.

\subsection{Quadruple Dataset Splits}
Referencing Section~\ref{rw: dataset}, conventional data splits (training, development, test) fail to precisely assess model performance on new queries, documents, or zero-shot scenarios. To address this, we adopt a multi-dimensional split strategy: splitting queries into an 80\% training and 20\% evaluation split, and splitting documents into two equal halves. This approach results in four sets: training, novel query, novel document, and zero-shot, with the latter comprising entirely unseen queries and documents. For evaluation, training and novel queries are tested against the first document corpus, while novel documents and zero-shot evaluations are conducted with the second corpus, ensuring a consistent corpus size across evaluations. The quadruple-split framework, depicted in Figure~\ref{fig: dp}, mirrors the varied challenges faced by real-world search systems, which encounter all four domains in differing volumes. 
In practical terms, evaluations conducted using the same data on which the model was trained are referred to as in-domain searches, while the other three evaluations represent various cold-start search scenarios.

\section{Experiments}
In this section, we evaluate the performance of GCL, firstly presenting ablation studies on score-to-weight functions and multi-Field weight $\gamma_{R_1}$ for documents. Following this, we compare its retrieval and ranking outcomes with the original CLIP~\cite{radford2021learning} and other public contrastive learning methods using MarqoGS-10M. GCL is fine-tuned from pre-trained models sourced from OpenClip~\cite{cherti2023reproducible}. To ensure a robust evaluation, we sampled 5000 queries for both development and test sets across all four evaluation splits. The dev queries are utilized for ablation studies. Comparisons with publicly available methods are conducted using the test queries.

\subsection{Evaluation Metrics}
We use three metrics to measure the ranked retrieval performance.

\paragraphseciton{Normalized Discounted Cumulative Gain (NDCG).}
NDCG~\cite{jarvelin2002cumulated} is one of the most widely used ranking measures for documents with graded relevance. For a single query, it is defined as $\text{NDCG}=\frac{\text{DCG}}{\text{IDCG}}$ where $\text{DCG}=\sum_{i=1}^{n_{doc}} \frac{s_i}{\log_2{(i+1)}}$, $n_{doc}$ is the number of documents for the query, and $\text{IDCG}$ is the ideal DCG which is DCG with the ground-truth ranking order.

\paragraphseciton{Expected Reciprocal Rank (ERR).}
ERR~\cite{chapelle2009expected} is an extension of the traditional reciprocal rank metric to incorporate graded relevance. For a query, it is defined as $\text{ERR} = \sum_{i=1}^{n_{doc}}\frac{1}{i}\prod_{j=1}^{i-1}\left(1-R(s_j)\right)R(s_i)$ where $R(\cdot)$ is a mapping function from graded relevance to probability defined as $R(s_i)=\frac{s_i}{s_{max}+1}$, and $s_{max}$ is the maximum relevance score for the query.

\paragraphseciton{Rank Based Precision(RBP).}
RBP~\cite{moffat2008rank} is a retrieval metric that models users' persistence of progressing from a document to the next document in a ranked list. For a single query, it is defined as $\text{RBP}=(1-p)\sum_{i=1}^{n_{doc}}\frac{s_i}{s_{max}}p^{i-1}$ where $p$ is a hyperparameter representing users' persistence. In our experiments, we fixed $p$ as 0.9, so that a user is expected looks at 10 items on average.


  







\renewcommand{\arraystretch}{1.35}
\begin{table}[t]
\caption{Comparing performance of GCL with various score-to-weight functions using GSFashion-5M and ViTB32~\cite{radford2021learning}.}
\vspace{0.5em}
\label{Tab: STW}
\small
\centering

\begin{tabular}{lccc|ccc}
\toprule
  & \multicolumn{3}{c|}{\textbf{In-Domain}} & \multicolumn{3}{c}{\textbf{Zero-Shot}}\\
 
 \textbf{STW} & \textbf{nDCG} & \textbf{ERR} & \textbf{RBP}  & \textbf{nDCG} & \textbf{ERR} & \textbf{RBP} \\

\hline

 Constant & 0.419 & 0.088 & 0.367 & 0.194 & 0.063 & 0.158  \\

 Linear & 0.583 & 0.163 & \textbf{0.483} & 0.201 & 0.073 & 0.163 \\

 Inverse & 0.599 & \textbf{0.608} & 0.459 & 0.201 & 0.090 & 0.165    \\
 Inverse Sqrt. & 0.561 & 0.322 & 0.456 & 0.198 & 0.077 & 0.161   \\

 Piecewise & \textbf{0.649} & 0.407 & 0.477 & \textbf{0.204} & \textbf{0.096} & \textbf{0.166} \\

~\vspace{-2em}
\end{tabular}
\end{table}

\renewcommand{\arraystretch}{1.35}
\begin{table}[h]
\caption{Performance comparison of GCL multi-field with varying image weight $\gamma_{R_1}$ using the GSFashion-5M dataset and VITB32~\cite{radford2021learning}. The last row represents the setting where a hybrid $\gamma_{R_1}$ equals 0.5 for in-domain and 0 for others.}
\label{Tab: mf}
\small
\centering

\begin{tabular}{cccc|ccc}
\toprule
  & \multicolumn{3}{c|}{\textbf{In-Domain}} & \multicolumn{3}{c}{\textbf{Zero-Shot}}\\
 
 \textbf{$\gamma_{R_1}$} & \textbf{nDCG} & \textbf{ERR} & \textbf{RBP}  & \textbf{nDCG} & \textbf{ERR} & \textbf{RBP} \\

\hline

 1.0 & 0.494 & 0.591 & 0.364 & 0.068 & 0.037 & 0.057  \\

 0.9 & 0.488 & 0.579 & 0.358 & 0.100 & 0.048 & 0.082  \\

 0.5 & \textbf{0.599} & \textbf{0.608} & \textbf{0.459} & 0.201 & 0.090 & 0.165  \\

 0.1 & 0.488 & 0.490 & 0.387 & 0.225 & \textbf{0.100} & 0.186  \\

 0.0 & 0.473 & 0.483 & 0.379 & \textbf{0.229} & 0.098 & \textbf{0.188}  \\

 \hline

 0.5/0.0 & \textbf{0.599} & \textbf{0.608} & \textbf{0.459} & \textbf{0.229} & 0.098 & \textbf{0.188}  
\end{tabular}
\end{table}

\renewcommand{\arraystretch}{1.35}
\begin{table*}[t]

\setlength{\tabcolsep}{5pt}
\caption{Retrieval and Ranking performance comparison of GCL versus publicly available contrastive learning methods~\cite{wang2022text, cherti2023reproducible, li2021learning, zhai2023sigmoid, robertson1995okapi, liu2019roberta} assessed by NDCG@10, ERR, and RBP metrics on MarqoGS-10M. Encoders denoted with "*" have pre-trained weights from original sources and OpenClip~\cite{cherti2023reproducible, wang2022text}.}

\label{Tab: Main results}
\small
\centering

\begin{tabular}{l|llccc|ccc|ccc|ccc}
\toprule

  & &  & \multicolumn{3}{c|}{\textbf{In-Domain}} & \multicolumn{3}{c|}{\textbf{Novel Queries}} & \multicolumn{3}{c|}{\textbf{Novel Corpus}}  & \multicolumn{3}{c}{\textbf{Zero-Shot}}\\
& \textbf{Methods} & \textbf{Encoders} & \textbf{nDCG} & \textbf{ERR} & \textbf{RBP}  & \textbf{nDCG} & \textbf{ERR} & \textbf{RBP} & \textbf{nDCG} & \textbf{ERR} & \textbf{RBP} & \textbf{nDCG} & \textbf{ERR} & \textbf{RBP} \\

\hline

\parbox[t]{2mm}{\multirow{7}{*}{\rotatebox[origin=c]{90}{\textbf{text-only}}}} 
& BM25~\cite{robertson1995okapi} & - & 0.071	    & 0.028 & 0.052 & 0.067 & 0.026 & 0.049 & 0.071 & 0.024 & 0.053 & 0.068  & 0.026 & 0.050\\

& Pretrained~\cite{wang2022text} & E5L~\cite{wang2022text} & 0.150 & 0.061 & 0.118 & 0.147 & 0.058 & 0.116 & 0.147 & 0.059 & 0.117 & 0.150 & 0.063 & 0.116     \\
& E5~\cite{wang2022text} & E5L~\cite{wang2022text} & 0.335 & 0.095 & 0.289 & 0.262 & 0.090 & 0.217 & 0.276 & 0.084 & 0.231 & 0.258 & 0.090 & 0.213     \\

& Pretrained~\cite{cherti2023reproducible} & RobB~\cite{liu2019roberta}      & 0.102 & 0.033 & 0.077 & 0.106 & 0.038 & 0.078 & 0.104 & 0.030 & 0.077 & 0.105 & 0.035 & 0.078\\
& Cross E. & RobB~\cite{liu2019roberta}     & 0.332 & 0.099 & 0.288 & 0.272 & 0.091 & 0.225 & 0.280 & 0.090 & 0.236 & 0.263 & 0.088 & 0.217\\
& GCL(ours) & E5L~\cite{wang2022text} & 0.431 & 0.400 & 0.347 & 0.299 & 0.172 & 0.244 & 0.286 & 0.119 & 0.239 & 0.271 & 0.116 & 0.223  \\
& GCL(ours) & RobB~\cite{liu2019roberta}   & \textbf{0.441} & \textbf{0.404} & \textbf{0.355} & \textbf{0.312} & \textbf{0.175} & \textbf{0.253} & \textbf{0.294} & \textbf{0.125} & \textbf{0.245} & \textbf{0.279} & \textbf{0.128} & \textbf{0.229}
 \\

\hline

\parbox[t]{2mm}{\multirow{8}{*}{\rotatebox[origin=c]{90}{\textbf{image-only}}}} 
& Pretrained~\cite{radford2021learning} & ViTB32~\cite{radford2021learning} & 0.063 & 0.025 & 0.052 & 0.063 & 0.024 & 0.052 & 0.061 & 0.020 & 0.051 & 0.063 & 0.024 & 0.052 \\
& CLIP~\cite{radford2021learning} & ViTB32~\cite{radford2021learning} & 0.258 & 0.059 & 0.228 & 0.096 & 0.032& 0.082 & 0.102 & 0.034 & 0.087 & 0.067 & 0.021 & 0.058\\


& Pretrained~\cite{radford2021learning} & ViTL14~\cite{radford2021learning} & 0.081 & 0.031 & 0.067 & 0.077 & 0.027 & 0.063 & 0.079 & 0.029 & 0.065 & 0.079 & 0.026 & 0.065 \\
& CLIP~\cite{radford2021learning} & ViTL14~\cite{radford2021learning} & 0.326 & 0.068 & 0.281 & 0.116 & 0.038 & 0.100 & \textbf{0.137} & 0.040 & \textbf{0.116} & 0.089 & 0.032 & 0.076 \\

& SigLip~\cite{zhai2023sigmoid} & ViTB16~\cite{radford2021learning} & 0.168 & 0.042 & 0.139 & 0.087 & 0.030 & 0.072 & 0.092 & 0.029 & 0.076 & 0.070 & 0.023 & 0.058 \\

& GCL(ours) & ViTB16~\cite{radford2021learning} & 0.234 & 0.172 & 0.176 & 0.159 & 0.122 & 0.123 & 0.125 & 0.046 & 0.103 & 0.071 & 0.026 & 0.058 \\

& GCL(ours)  & ViTB32~\cite{radford2021learning} & 0.449 & \textbf{0.564} & 0.329 & 0.141 & 0.124 & 0.111 & 0.101 & 0.040 & 0.086 & 0.074 & 0.032 & 0.062 \\
& GCL(ours)  & ViTL14~\cite{radford2021learning} & \textbf{0.489} & 0.530 & \textbf{0.362} & \textbf{0.160} & \textbf{0.124} & \textbf{0.127} & 0.125 & \textbf{0.047} & 0.104 & \textbf{0.091} & \textbf{0.036} & \textbf{0.078} \\

\hline
\parbox[t]{2mm}{\multirow{3}{*}{\rotatebox[origin=c]{90}{\textbf{multi}}}}
& CLIP & ViTL14~\cite{radford2021learning} & 0.310 & 0.093 & 0.252 & 0.205 & 0.075 & 0.165 & 0.228 & 0.081 & 0.184 & 0.199 & 0.079 & 0.159 \\
& GCL(ours)  & ViTB32~\cite{radford2021learning} & 0.577 & 0.554 & 0.446 & 0.287 & 0.144 & 0.237 & 0.276 & 0.110 & 0.231 & 0.257 & 0.108 & 0.213 \\ 
& GCL(ours)  & ViTL14~\cite{radford2021learning} & \textbf{0.603} & \textbf{0.562} & \textbf{0.467} & \textbf{0.305} & \textbf{0.156} & \textbf{0.251} & \textbf{0.288} & \textbf{0.118} & \textbf{0.241} & \textbf{0.272} & \textbf{0.114} & \textbf{0.224} \\ 
\end{tabular}

\end{table*}

\subsection{Ablation studies}
\paragraphseciton{Score-to-Weight Function.} We evaluate five distinct STW functions discussed in Sec~\ref{sec: weighted-clip} and results are shown in Table~\ref{Tab: STW}.
The Constant function serves as our baseline, reflecting the unweighted loss approach common in conventional contrastive learning \cite{li2021learning}. In contrast, the Linear function, which directly applies the ranking scores as weights, demonstrated notable enhancements over the baseline in all tested scenarios.
The Inverse function, adjusting weight distribution to prioritize pairs with higher scores, shows improvement across all metrics relative to the Linear approach. The performance of inverse function for NDCG has gained \textbf{18.0\%} for In-domain and \textbf{0.7\%} for the zero-shot compared to the baseline, highlighting its strength in prioritizing pairs with top scores. 
The Piecewise function is designed to assign equal weights to the top 10\% of documents for a given query, aligning well with the NDCG@10 metric. Consequently, it has achieved significant improvements in NDCG@10, showing a \textbf{23.0\%} increase for In-domain and \textbf{1.0\%} for zero-shot compared to the baseline.
This illustrates how GCL can be tailored to focus on different metrics in a practical setting. We use inverse for the rest of experiments.

\paragraphseciton{Multi-Field weight $\gamma_{R_1}$ for document.} In this analysis, we use product image and title as RHS fields to represent the document. In calculating $\mathbf{Z}_{avg}$, discussed in Section~\ref{subsec: multi-field}, we assign weight $\gamma{R_1}$ to the image field and $\gamma_{R_2}$ to the title field, with $\gamma_{R_2} = 1 - \gamma_{R_1}$. 

The result in Table~\ref{Tab: mf} shows that the model performs the best in in-domain evaluation with $\gamma_{R_1}$ equals to 0.5, signifying an even 50/50 image and title contribution to the average embedding. Conversely, for zero-shot evaluation, the model exhibits optimal results when relying solely on the title field. The integration of pairwise loss between each field on the LHS and each field on the RHS enables the use of pure title data, even though the image data is also trained as part of the RHS fields. Therefore, we conducted additional evaluations setting $\gamma_{R_1}$ to 0.5 for in-domain and to 0 for zero-shot evaluation, yielding the most favorable outcomes overall. We thus use the 0.5/0.0 setting for the multi-field experiment in Table~\ref{Tab: Main results}.

\vspace{-1em}
\subsection{Comparison with Public Contrastive Learning Methods}
\label{subsec: compare main}

This subsection presents a comparative analysis of the retrieval and ranking performance of GCL against established public contrastive learning frameworks. We conduct evaluations across four data splits on MarqoGS-10M dataset in Table~\ref{Tab: Main results}. The evaluation considers text-based queries against three document formats: text-only (title), image-only, and multi-field (title and image). To maintain a fair comparison, all finetuned models listed in this table have been trained over 20 epochs with a batch size of 2048. 

\paragraphseciton{Text-To-Text Retrieval.}
In this study, we represent documents solely by product titles, thereby constructing a text-to-text retrieval scenario. We utilize the widely adopted BM25~\cite{robertson1995okapi} for initial comparisons. Our findings reveal that pre-trained text models, specifically E5~\cite{wang2022text} and Roberta~\cite{robertson1995okapi}, surpass BM25's performance. Further enhancements are observed when these models are fine-tuned on the MarqoGS-10M dataset using their original frameworks. Traditional frameworks cannot utilize rankings. In contrast, models fine-tuned using our GCL approach significantly outperform conventional methods in in-domain evaluations and exhibit marked improvements in cold-start scenarios. Numerically, our top-performing model, RobB, fine-tuned with GCL, has achieved a \textbf{10.9\%} increase in NDCG@10 and a \textbf{30.5\%} increase in ERR compared against RobB fine-tuned using conventional cross entropy loss for in-domain evaluation. For cold-start scenarios, it shows a \textbf{1.4 - 5\%} improvement in NDCG@10 and a \textbf{3.5 - 8.4\%} enhancement in ERR.

\paragraphseciton{Text-To-Image Retrieval.} In this analysis, we represent documents solely by product thumbnail images, thereby constructing a text-to-image retrieval scenario. Results demonstrate that models fine-tuned with CLIP and SigLip on the GSFull-10M dataset outperform their pre-trained counterparts. Yet, these conventional approaches do not incorporate ranking scores. Conversely, our GCL framework, when applied for fine-tuning, markedly exceeds the performance of standard methods in in-domain assessments and shows substantial enhancements in cold-start conditions. Numerically, our best-performing model, VITL14, fine-tuned with GCL, realized a \textbf{16.3\%} increase in NDCG@10 and a \textbf{46.2\%} increase in ERR compared to VITL14 fine-tuned using the original CLIP method~\cite{radford2021learning} for in-domain assessments. For cold-start situations, it exhibited an enhancement of \textbf{0.4\% - 8.6\%} in ERR. The exception occurs in the NDCG@10 metric for novel corpus evaluations, where GCL slightly underperforms baseline. Nonetheless, GCL exceeds CLIP in NDCG@10 by \textbf{4.4\%} and \textbf{0.2\%} in Novel Queries and Zero-Shot.

\paragraphseciton{Multi-Field Retrieval.} The multi-field implementation of our GCL framework, which represents documents using both product images and titles, has significantly outperformed text-only and image-only counterparts, delivering the best overall results. The optimal configurations identified in the ablation studies were applied. While the original CLIP~\cite{radford2021learning} also shows enhanced performance, it does not employ multi-field training. GCL's integration of multi-field data and ranking signals has led to a remarkable \textbf{29.3\%} increase in NDCG@10 and a \textbf{46.9\%} increase in ERR compared to VITL14 fine-tuned with cross entropy loss in the original CLIP method. In cold-start scenarios, GCL has shown superior performance to CLIP~\cite{radford2021learning}, with improvements of \textbf{6.0\% - 10.0\%} in NDCG@10, \textbf{3.5\% - 8.1\%} in ERR, and \textbf{5.7\% - 8.6\%} in RBP. 

\renewcommand{\arraystretch}{1.35}
\begin{table*}[t]

\setlength{\tabcolsep}{5pt}
\caption{Retrieval and Ranking performance comparison of GCL against baseline on proprietary ecommerce data with.}

\label{Tab: Offline}
\small
\centering

\begin{tabular}{llccc|ccc|ccc|ccc}
\toprule

  &  & \multicolumn{3}{c|}{\textbf{In-Domain}} & \multicolumn{3}{c|}{\textbf{Novel Queries}} & \multicolumn{3}{c|}{\textbf{Novel Corpus}}  & \multicolumn{3}{c}{\textbf{Zero-Shot}}\\
 \textbf{Methods} & \textbf{Encoders} & \textbf{nDCG} & \textbf{ERR} & \textbf{RBP}  & \textbf{nDCG} & \textbf{ERR} & \textbf{RBP} & \textbf{nDCG} & \textbf{ERR} & \textbf{RBP} & \textbf{nDCG} & \textbf{ERR} & \textbf{RBP} \\

\hline

Pretrained & VITB32~\cite{radford2021learning}  & 0.245 & 0.155 & 0.102 & 0.249 & 0.161 & 0.100 & 0.285 & 0.195 & 0.107 & 0.274 & 0.181 & 0.100 \\
Cross E. & VITB32~\cite{radford2021learning}  & 0.447 & 0.256 & 0.185 & 0.319 & 0.215 & 0.125 & 0.296 & 0.187 & 0.123 & \textbf{0.303} & \textbf{0.202} & \textbf{0.114}\\
GCL(ours). & VITB32~\cite{radford2021learning}  & \textbf{0.501} & \textbf{0.316} & \textbf{0.222} & \textbf{0.327} & \textbf{0.222} & \textbf{0.127} & \textbf{0.326} & \textbf{0.221} & \textbf{0.127} & 0.298 & 0.198 & 0.113 \\

\hline
Pretrained & VITL14~\cite{radford2021learning}  & 0.246 & 0.156 & 0.100 & 0.252 & 0.163 & 0.099 & 0.287 & \textbf{0.195} & 0.106 & 0.272 & 0.183 & 0.099 \\
Cross E. & VITL14~\cite{radford2021learning}  & 0.473 & 0.265 & 0.198 & \textbf{0.338} & \textbf{0.228} & \textbf{0.129} & \textbf{0.306} & 0.194 & \textbf{0.126} & \textbf{0.312} & \textbf{0.207} & \textbf{0.117} \\
GCL(ours). & VITL14~\cite{radford2021learning}  & \textbf{0.585} & \textbf{0.383} & \textbf{0.260} & 0.335 & \textbf{0.228} & \textbf{0.129} & 0.300 & 0.190 & \textbf{0.126} & 0.304 & 0.200 & 0.115 \\

\end{tabular}

\end{table*}


\begin{figure}
\includegraphics[width=0.9\linewidth]{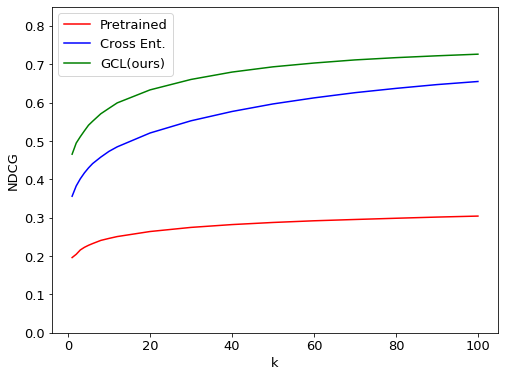}
\caption{NDCG vs k over proprietary in-domain data. }
\vspace{-2em}
\label{fig: ndcg_k}
\end{figure}

\vspace{-1em}
\subsection{Offline results on Ecommerce Data}
In this experiment, we evaluate the effectiveness of GCL using user interaction data collected from an e-commerce platform. Specifically, we record the number of times users added a product to the cart (ATC) after a search query. Each data entry includes the search query, product title, product image, and the ATC count, which ranges from 1 to 2.6k, with a mean of 2.28 and a median of 1. To moderate the effect of ATC in evaluation and training, we compute a score: $s = \log_{1.1}(ATC) + 1$,  resulting in values that approximately range from 1 to 100, aligning with the scale of our marqo-gs-10m dataset. We split this data as outlined in Figure~\ref{fig: dp}. All training and evaluation settings mirror those in Section~\ref{subsec: compare main}, with a linear score-to-weight function applied. 
Results, shown in Table~\ref{Tab: Offline}, demonstrate that models fine-tuned with GCL achieve substantial improvements for in-domain evaluation. Notably, we observe up to an \textbf{11.2\%} increase in nDCG@10 compared to the baseline model fine-tuned with conventional cross-entropy loss. As illustrated in Figure~\ref{fig: ndcg_k}, GCL consistently outperforms the baseline across all $k$-values in nDCG. In cold-start evaluations, GCL performs comparably to the baseline, with only a minor 0.8\% decrease in nDCG for zero-shot scenarios, where both queries and products are unseen in the training set.
Interestingly, despite being a smaller model, VITB32 outperforms VITL14 on several cold-start metrics. We attribute this to potential overfitting in VITL14 given the current parameter settings for this specific dataset, though exploring this further is beyond the scope. Importantly, as over 90\% of traffic in real-world applications is in-domain, the VITL14 model trained with GCL has delivered a notable 3.2\% increase in revenue in an A/B test against the baseline.

\vspace{-1em}
\section{Conclusion}
In this work, we show that the current contrastive learning approaches fail to incorporate direct ranking signal leading to low ranked retrieval NDCG performance. To address the gap, we first acquire a large-scale dataset MarqoGS-10M with ranking scores to support the research community. Following that, we propose GCL, which integrates ranking information and train with multi-field query and document. GCL has surpassed conventional methods by a large margin in both MarqoGS-10M and an offline evaluation with proprietary ecommerce data, indicating significant potential for future research. One limitation of GCL is that it slows the learning of relevant documents with low ranking scores, which consequently reduces recall at higher values of $k$. Addressing this issue is a subject for future research.

\bibliographystyle{ACM-Reference-Format}
\balance
\bibliography{acmart}


\begin{thebibliography}{74}


\ifx \showCODEN    \undefined \def \showCODEN     #1{\unskip}     \fi
\ifx \showDOI      \undefined \def \showDOI       #1{#1}\fi
\ifx \showISBNx    \undefined \def \showISBNx     #1{\unskip}     \fi
\ifx \showISBNxiii \undefined \def \showISBNxiii  #1{\unskip}     \fi
\ifx \showISSN     \undefined \def \showISSN      #1{\unskip}     \fi
\ifx \showLCCN     \undefined \def \showLCCN      #1{\unskip}     \fi
\ifx \shownote     \undefined \def \shownote      #1{#1}          \fi
\ifx \showarticletitle \undefined \def \showarticletitle #1{#1}   \fi
\ifx \showURL      \undefined \def \showURL       {\relax}        \fi
\providecommand\bibfield[2]{#2}
\providecommand\bibinfo[2]{#2}
\providecommand\natexlab[1]{#1}
\providecommand\showeprint[2][]{arXiv:#2}

\bibitem[Achiam et~al\mbox{.}(2023)]%
        {achiam2023gpt}
\bibfield{author}{\bibinfo{person}{Josh Achiam}, \bibinfo{person}{Steven Adler}, \bibinfo{person}{Sandhini Agarwal}, \bibinfo{person}{Lama Ahmad}, \bibinfo{person}{Ilge Akkaya}, \bibinfo{person}{Florencia~Leoni Aleman}, \bibinfo{person}{Diogo Almeida}, \bibinfo{person}{Janko Altenschmidt}, \bibinfo{person}{Sam Altman}, \bibinfo{person}{Shyamal Anadkat}, {et~al\mbox{.}}} \bibinfo{year}{2023}\natexlab{}.
\newblock \showarticletitle{Gpt-4 technical report}.
\newblock \bibinfo{journal}{\emph{arXiv preprint arXiv:2303.08774}} (\bibinfo{year}{2023}).
\newblock


\bibitem[Asadi and Lin(2013)]%
        {asadi2013effectiveness}
\bibfield{author}{\bibinfo{person}{Nima Asadi} {and} \bibinfo{person}{Jimmy Lin}.} \bibinfo{year}{2013}\natexlab{}.
\newblock \showarticletitle{Effectiveness/efficiency tradeoffs for candidate generation in multi-stage retrieval architectures}. In \bibinfo{booktitle}{\emph{Proceedings of the 36th international ACM SIGIR conference on Research and development in information retrieval}}. \bibinfo{pages}{997--1000}.
\newblock


\bibitem[Bondarenko et~al\mbox{.}(2020)]%
        {bondarenko2020overview}
\bibfield{author}{\bibinfo{person}{Alexander Bondarenko}, \bibinfo{person}{Maik Fr{\"o}be}, \bibinfo{person}{Meriem Beloucif}, \bibinfo{person}{Lukas Gienapp}, \bibinfo{person}{Yamen Ajjour}, \bibinfo{person}{Alexander Panchenko}, \bibinfo{person}{Chris Biemann}, \bibinfo{person}{Benno Stein}, \bibinfo{person}{Henning Wachsmuth}, \bibinfo{person}{Martin Potthast}, {et~al\mbox{.}}} \bibinfo{year}{2020}\natexlab{}.
\newblock \showarticletitle{Overview of Touch{\'e} 2020: argument retrieval}. In \bibinfo{booktitle}{\emph{Experimental IR Meets Multilinguality, Multimodality, and Interaction: 11th International Conference of the CLEF Association, CLEF 2020, Thessaloniki, Greece, September 22--25, 2020, Proceedings 11}}. Springer, \bibinfo{pages}{384--395}.
\newblock


\bibitem[Bonifacio et~al\mbox{.}(2022)]%
        {bonifacio2022inpars}
\bibfield{author}{\bibinfo{person}{Luiz Bonifacio}, \bibinfo{person}{Hugo Abonizio}, \bibinfo{person}{Marzieh Fadaee}, {and} \bibinfo{person}{Rodrigo Nogueira}.} \bibinfo{year}{2022}\natexlab{}.
\newblock \showarticletitle{Inpars: Data augmentation for information retrieval using large language models}.
\newblock \bibinfo{journal}{\emph{arXiv preprint arXiv:2202.05144}} (\bibinfo{year}{2022}).
\newblock


\bibitem[Boteva et~al\mbox{.}(2016)]%
        {boteva2016full}
\bibfield{author}{\bibinfo{person}{Vera Boteva}, \bibinfo{person}{Demian Gholipour}, \bibinfo{person}{Artem Sokolov}, {and} \bibinfo{person}{Stefan Riezler}.} \bibinfo{year}{2016}\natexlab{}.
\newblock \showarticletitle{A full-text learning to rank dataset for medical information retrieval}. In \bibinfo{booktitle}{\emph{Advances in Information Retrieval: 38th European Conference on IR Research, ECIR 2016, Padua, Italy, March 20--23, 2016. Proceedings 38}}. Springer, \bibinfo{pages}{716--722}.
\newblock


\bibitem[Changpinyo et~al\mbox{.}(2021)]%
        {changpinyo2021conceptual}
\bibfield{author}{\bibinfo{person}{Soravit Changpinyo}, \bibinfo{person}{Piyush Sharma}, \bibinfo{person}{Nan Ding}, {and} \bibinfo{person}{Radu Soricut}.} \bibinfo{year}{2021}\natexlab{}.
\newblock \showarticletitle{Conceptual 12m: Pushing web-scale image-text pre-training to recognize long-tail visual concepts}. In \bibinfo{booktitle}{\emph{Proceedings of the IEEE/CVF conference on computer vision and pattern recognition}}. \bibinfo{pages}{3558--3568}.
\newblock


\bibitem[Chapelle et~al\mbox{.}(2009)]%
        {chapelle2009expected}
\bibfield{author}{\bibinfo{person}{Olivier Chapelle}, \bibinfo{person}{Donald Metlzer}, \bibinfo{person}{Ya Zhang}, {and} \bibinfo{person}{Pierre Grinspan}.} \bibinfo{year}{2009}\natexlab{}.
\newblock \showarticletitle{Expected reciprocal rank for graded relevance}. In \bibinfo{booktitle}{\emph{Proceedings of the 18th ACM Conference on Information and Knowledge Management}} (Hong Kong, China) \emph{(\bibinfo{series}{CIKM '09})}. \bibinfo{publisher}{Association for Computing Machinery}, \bibinfo{address}{New York, NY, USA}, \bibinfo{pages}{621–630}.
\newblock
\showISBNx{9781605585123}
\urldef\tempurl%
\url{https://doi.org/10.1145/1645953.1646033}
\showDOI{\tempurl}


\bibitem[Chen et~al\mbox{.}(2020)]%
        {chen2020simple}
\bibfield{author}{\bibinfo{person}{Ting Chen}, \bibinfo{person}{Simon Kornblith}, \bibinfo{person}{Mohammad Norouzi}, {and} \bibinfo{person}{Geoffrey Hinton}.} \bibinfo{year}{2020}\natexlab{}.
\newblock \showarticletitle{A Simple Framework for Contrastive Learning of Visual Representations}. In \bibinfo{booktitle}{\emph{Proceedings of the 37th International Conference on Machine Learning}} \emph{(\bibinfo{series}{Proceedings of Machine Learning Research}, Vol.~\bibinfo{volume}{119})}, \bibfield{editor}{\bibinfo{person}{Hal~Daumé III} {and} \bibinfo{person}{Aarti Singh}} (Eds.). \bibinfo{publisher}{PMLR}, \bibinfo{pages}{1597--1607}.
\newblock
\urldef\tempurl%
\url{https://proceedings.mlr.press/v119/chen20j.html}
\showURL{%
\tempurl}


\bibitem[Cherti et~al\mbox{.}(2023)]%
        {cherti2023reproducible}
\bibfield{author}{\bibinfo{person}{Mehdi Cherti}, \bibinfo{person}{Romain Beaumont}, \bibinfo{person}{Ross Wightman}, \bibinfo{person}{Mitchell Wortsman}, \bibinfo{person}{Gabriel Ilharco}, \bibinfo{person}{Cade Gordon}, \bibinfo{person}{Christoph Schuhmann}, \bibinfo{person}{Ludwig Schmidt}, {and} \bibinfo{person}{Jenia Jitsev}.} \bibinfo{year}{2023}\natexlab{}.
\newblock \showarticletitle{Reproducible Scaling Laws for Contrastive Language-Image Learning}. In \bibinfo{booktitle}{\emph{Proceedings of the IEEE/CVF Conference on Computer Vision and Pattern Recognition (CVPR)}}. \bibinfo{pages}{2818--2829}.
\newblock


\bibitem[Craswell et~al\mbox{.}(2020)]%
        {craswell2020overview}
\bibfield{author}{\bibinfo{person}{Nick Craswell}, \bibinfo{person}{Bhaskar Mitra}, \bibinfo{person}{Emine Yilmaz}, \bibinfo{person}{Daniel Campos}, {and} \bibinfo{person}{Ellen~M Voorhees}.} \bibinfo{year}{2020}\natexlab{}.
\newblock \showarticletitle{Overview of the TREC 2019 deep learning track}.
\newblock \bibinfo{journal}{\emph{arXiv preprint arXiv:2003.07820}} (\bibinfo{year}{2020}).
\newblock


\bibitem[Deng et~al\mbox{.}(2009)]%
        {deng2009imagenet}
\bibfield{author}{\bibinfo{person}{Jia Deng}, \bibinfo{person}{Wei Dong}, \bibinfo{person}{Richard Socher}, \bibinfo{person}{Li-Jia Li}, \bibinfo{person}{Kai Li}, {and} \bibinfo{person}{Li Fei-Fei}.} \bibinfo{year}{2009}\natexlab{}.
\newblock \showarticletitle{Imagenet: A large-scale hierarchical image database}. In \bibinfo{booktitle}{\emph{2009 IEEE conference on computer vision and pattern recognition}}. Ieee, \bibinfo{pages}{248--255}.
\newblock


\bibitem[Desai et~al\mbox{.}(2021)]%
        {desai2021redcaps}
\bibfield{author}{\bibinfo{person}{Karan Desai}, \bibinfo{person}{Gaurav Kaul}, \bibinfo{person}{Zubin Aysola}, {and} \bibinfo{person}{Justin Johnson}.} \bibinfo{year}{2021}\natexlab{}.
\newblock \showarticletitle{Redcaps: Web-curated image-text data created by the people, for the people}.
\newblock \bibinfo{journal}{\emph{arXiv preprint arXiv:2111.11431}} (\bibinfo{year}{2021}).
\newblock


\bibitem[Diggelmann et~al\mbox{.}(2020)]%
        {diggelmann2020climate}
\bibfield{author}{\bibinfo{person}{Thomas Diggelmann}, \bibinfo{person}{Jordan Boyd-Graber}, \bibinfo{person}{Jannis Bulian}, \bibinfo{person}{Massimiliano Ciaramita}, {and} \bibinfo{person}{Markus Leippold}.} \bibinfo{year}{2020}\natexlab{}.
\newblock \showarticletitle{Climate-fever: A dataset for verification of real-world climate claims}.
\newblock \bibinfo{journal}{\emph{arXiv preprint arXiv:2012.00614}} (\bibinfo{year}{2020}).
\newblock


\bibitem[Elizalde et~al\mbox{.}(2023)]%
        {elizalde2023clap}
\bibfield{author}{\bibinfo{person}{Benjamin Elizalde}, \bibinfo{person}{Soham Deshmukh}, \bibinfo{person}{Mahmoud Al~Ismail}, {and} \bibinfo{person}{Huaming Wang}.} \bibinfo{year}{2023}\natexlab{}.
\newblock \showarticletitle{Clap learning audio concepts from natural language supervision}. In \bibinfo{booktitle}{\emph{ICASSP 2023-2023 IEEE International Conference on Acoustics, Speech and Signal Processing (ICASSP)}}. IEEE, \bibinfo{pages}{1--5}.
\newblock


\bibitem[Frej et~al\mbox{.}(2020)]%
        {frej2020learning}
\bibfield{author}{\bibinfo{person}{Jibril Frej}, \bibinfo{person}{Philippe Mulhem}, \bibinfo{person}{Didier Schwab}, {and} \bibinfo{person}{Jean-Pierre Chevallet}.} \bibinfo{year}{2020}\natexlab{}.
\newblock \showarticletitle{Learning Term Discrimination}. In \bibinfo{booktitle}{\emph{Proceedings of the 43rd International ACM SIGIR Conference on Research and Development in Information Retrieval}} (Virtual Event, China) \emph{(\bibinfo{series}{SIGIR '20})}. \bibinfo{publisher}{Association for Computing Machinery}, \bibinfo{address}{New York, NY, USA}, \bibinfo{pages}{1993–1996}.
\newblock
\showISBNx{9781450380164}
\urldef\tempurl%
\url{https://doi.org/10.1145/3397271.3401211}
\showDOI{\tempurl}


\bibitem[Gao et~al\mbox{.}(2010)]%
        {gao2010clickthrough}
\bibfield{author}{\bibinfo{person}{Jianfeng Gao}, \bibinfo{person}{Xiaodong He}, {and} \bibinfo{person}{Jian-Yun Nie}.} \bibinfo{year}{2010}\natexlab{}.
\newblock \showarticletitle{Clickthrough-based translation models for web search: from word models to phrase models}. In \bibinfo{booktitle}{\emph{Proceedings of the 19th ACM international conference on Information and knowledge management}}. \bibinfo{pages}{1139--1148}.
\newblock


\bibitem[Gao et~al\mbox{.}(2011)]%
        {gao2011clickthrough}
\bibfield{author}{\bibinfo{person}{Jianfeng Gao}, \bibinfo{person}{Kristina Toutanova}, {and} \bibinfo{person}{Wen-tau Yih}.} \bibinfo{year}{2011}\natexlab{}.
\newblock \showarticletitle{Clickthrough-based latent semantic models for web search}. In \bibinfo{booktitle}{\emph{Proceedings of the 34th international ACM SIGIR conference on Research and development in Information Retrieval}}. \bibinfo{pages}{675--684}.
\newblock


\bibitem[Goodfellow et~al\mbox{.}(2020)]%
        {goodfellow2020generative}
\bibfield{author}{\bibinfo{person}{Ian Goodfellow}, \bibinfo{person}{Jean Pouget-Abadie}, \bibinfo{person}{Mehdi Mirza}, \bibinfo{person}{Bing Xu}, \bibinfo{person}{David Warde-Farley}, \bibinfo{person}{Sherjil Ozair}, \bibinfo{person}{Aaron Courville}, {and} \bibinfo{person}{Yoshua Bengio}.} \bibinfo{year}{2020}\natexlab{}.
\newblock \showarticletitle{Generative adversarial networks}.
\newblock \bibinfo{journal}{\emph{Commun. ACM}} \bibinfo{volume}{63}, \bibinfo{number}{11} (\bibinfo{year}{2020}), \bibinfo{pages}{139--144}.
\newblock


\bibitem[Gutmann and Hyvärinen(2010)]%
        {gutmann2010noise}
\bibfield{author}{\bibinfo{person}{Michael Gutmann} {and} \bibinfo{person}{Aapo Hyvärinen}.} \bibinfo{year}{2010}\natexlab{}.
\newblock \showarticletitle{Noise-contrastive estimation: A new estimation principle for unnormalized statistical models}. In \bibinfo{booktitle}{\emph{Proceedings of the Thirteenth International Conference on Artificial Intelligence and Statistics}} \emph{(\bibinfo{series}{Proceedings of Machine Learning Research}, Vol.~\bibinfo{volume}{9})}, \bibfield{editor}{\bibinfo{person}{Yee~Whye Teh} {and} \bibinfo{person}{Mike Titterington}} (Eds.). \bibinfo{publisher}{PMLR}, \bibinfo{address}{Chia Laguna Resort, Sardinia, Italy}, \bibinfo{pages}{297--304}.
\newblock
\urldef\tempurl%
\url{https://proceedings.mlr.press/v9/gutmann10a.html}
\showURL{%
\tempurl}


\bibitem[He et~al\mbox{.}(2020)]%
        {he2020momentum}
\bibfield{author}{\bibinfo{person}{Kaiming He}, \bibinfo{person}{Haoqi Fan}, \bibinfo{person}{Yuxin Wu}, \bibinfo{person}{Saining Xie}, {and} \bibinfo{person}{Ross Girshick}.} \bibinfo{year}{2020}\natexlab{}.
\newblock \showarticletitle{Momentum Contrast for Unsupervised Visual Representation Learning}. In \bibinfo{booktitle}{\emph{Proceedings of the IEEE/CVF Conference on Computer Vision and Pattern Recognition (CVPR)}}.
\newblock


\bibitem[Huang et~al\mbox{.}(2013)]%
        {huang2013learning}
\bibfield{author}{\bibinfo{person}{Po-Sen Huang}, \bibinfo{person}{Xiaodong He}, \bibinfo{person}{Jianfeng Gao}, \bibinfo{person}{Li Deng}, \bibinfo{person}{Alex Acero}, {and} \bibinfo{person}{Larry Heck}.} \bibinfo{year}{2013}\natexlab{}.
\newblock \showarticletitle{Learning deep structured semantic models for web search using clickthrough data}. In \bibinfo{booktitle}{\emph{Proceedings of the 22nd ACM international conference on Information \& Knowledge Management}}. \bibinfo{pages}{2333--2338}.
\newblock


\bibitem[Ioffe(2010)]%
        {ioffe2010improved}
\bibfield{author}{\bibinfo{person}{Sergey Ioffe}.} \bibinfo{year}{2010}\natexlab{}.
\newblock \showarticletitle{Improved consistent sampling, weighted minhash and l1 sketching}. In \bibinfo{booktitle}{\emph{2010 IEEE international conference on data mining}}. IEEE, \bibinfo{pages}{246--255}.
\newblock


\bibitem[Jaiswal et~al\mbox{.}(2021)]%
        {jaiswal2021survey}
\bibfield{author}{\bibinfo{person}{Ashish Jaiswal}, \bibinfo{person}{Ashwin~Ramesh Babu}, \bibinfo{person}{Mohammad~Zaki Zadeh}, \bibinfo{person}{Debapriya Banerjee}, {and} \bibinfo{person}{Fillia Makedon}.} \bibinfo{year}{2021}\natexlab{}.
\newblock \showarticletitle{A Survey on Contrastive Self-Supervised Learning}.
\newblock \bibinfo{journal}{\emph{Technologies}} \bibinfo{volume}{9}, \bibinfo{number}{1} (\bibinfo{year}{2021}).
\newblock
\showISSN{2227-7080}
\urldef\tempurl%
\url{https://doi.org/10.3390/technologies9010002}
\showDOI{\tempurl}


\bibitem[Jang et~al\mbox{.}(2021)]%
        {jang2021ultra}
\bibfield{author}{\bibinfo{person}{Kyoung-Rok Jang}, \bibinfo{person}{Junmo Kang}, \bibinfo{person}{Giwon Hong}, \bibinfo{person}{Sung-Hyon Myaeng}, \bibinfo{person}{Joohee Park}, \bibinfo{person}{Taewon Yoon}, {and} \bibinfo{person}{Heecheol Seo}.} \bibinfo{year}{2021}\natexlab{}.
\newblock \showarticletitle{Ultra-high dimensional sparse representations with binarization for efficient text retrieval}.
\newblock \bibinfo{journal}{\emph{arXiv preprint arXiv:2104.07198}} (\bibinfo{year}{2021}).
\newblock


\bibitem[J\"{a}rvelin and Kek\"{a}l\"{a}inen(2002)]%
        {jarvelin2002cumulated}
\bibfield{author}{\bibinfo{person}{Kalervo J\"{a}rvelin} {and} \bibinfo{person}{Jaana Kek\"{a}l\"{a}inen}.} \bibinfo{year}{2002}\natexlab{}.
\newblock \showarticletitle{Cumulated gain-based evaluation of IR techniques}.
\newblock \bibinfo{journal}{\emph{ACM Trans. Inf. Syst.}} \bibinfo{volume}{20}, \bibinfo{number}{4} (\bibinfo{date}{oct} \bibinfo{year}{2002}), \bibinfo{pages}{422–446}.
\newblock
\showISSN{1046-8188}
\urldef\tempurl%
\url{https://doi.org/10.1145/582415.582418}
\showDOI{\tempurl}


\bibitem[Jeronymo et~al\mbox{.}(2023)]%
        {jeronymo2023inpars}
\bibfield{author}{\bibinfo{person}{Vitor Jeronymo}, \bibinfo{person}{Luiz Bonifacio}, \bibinfo{person}{Hugo Abonizio}, \bibinfo{person}{Marzieh Fadaee}, \bibinfo{person}{Roberto Lotufo}, \bibinfo{person}{Jakub Zavrel}, {and} \bibinfo{person}{Rodrigo Nogueira}.} \bibinfo{year}{2023}\natexlab{}.
\newblock \showarticletitle{InPars-v2: Large Language Models as Efficient Dataset Generators for Information Retrieval}.
\newblock \bibinfo{journal}{\emph{arXiv preprint arXiv:2301.01820}} (\bibinfo{year}{2023}).
\newblock


\bibitem[Jia et~al\mbox{.}(2021)]%
        {jia2021scaling}
\bibfield{author}{\bibinfo{person}{Chao Jia}, \bibinfo{person}{Yinfei Yang}, \bibinfo{person}{Ye Xia}, \bibinfo{person}{Yi-Ting Chen}, \bibinfo{person}{Zarana Parekh}, \bibinfo{person}{Hieu Pham}, \bibinfo{person}{Quoc Le}, \bibinfo{person}{Yun-Hsuan Sung}, \bibinfo{person}{Zhen Li}, {and} \bibinfo{person}{Tom Duerig}.} \bibinfo{year}{2021}\natexlab{}.
\newblock \showarticletitle{Scaling Up Visual and Vision-Language Representation Learning With Noisy Text Supervision}. In \bibinfo{booktitle}{\emph{Proceedings of the 38th International Conference on Machine Learning}} \emph{(\bibinfo{series}{Proceedings of Machine Learning Research}, Vol.~\bibinfo{volume}{139})}, \bibfield{editor}{\bibinfo{person}{Marina Meila} {and} \bibinfo{person}{Tong Zhang}} (Eds.). \bibinfo{publisher}{PMLR}, \bibinfo{pages}{4904--4916}.
\newblock
\urldef\tempurl%
\url{https://proceedings.mlr.press/v139/jia21b.html}
\showURL{%
\tempurl}


\bibitem[Kalantidis et~al\mbox{.}(2020)]%
        {kalantidis2020hard}
\bibfield{author}{\bibinfo{person}{Yannis Kalantidis}, \bibinfo{person}{Mert~Bulent Sariyildiz}, \bibinfo{person}{Noe Pion}, \bibinfo{person}{Philippe Weinzaepfel}, {and} \bibinfo{person}{Diane Larlus}.} \bibinfo{year}{2020}\natexlab{}.
\newblock \showarticletitle{Hard negative mixing for contrastive learning}.
\newblock \bibinfo{journal}{\emph{Advances in neural information processing systems}}  \bibinfo{volume}{33} (\bibinfo{year}{2020}), \bibinfo{pages}{21798--21809}.
\newblock


\bibitem[Kenton and Toutanova(2019)]%
        {kenton2019bert}
\bibfield{author}{\bibinfo{person}{Jacob Devlin Ming-Wei~Chang Kenton} {and} \bibinfo{person}{Lee~Kristina Toutanova}.} \bibinfo{year}{2019}\natexlab{}.
\newblock \showarticletitle{Bert: Pre-training of deep bidirectional transformers for language understanding}. In \bibinfo{booktitle}{\emph{Proceedings of naacL-HLT}}, Vol.~\bibinfo{volume}{1}. \bibinfo{pages}{2}.
\newblock


\bibitem[Khattab and Zaharia(2020)]%
        {khattab2020colbert}
\bibfield{author}{\bibinfo{person}{Omar Khattab} {and} \bibinfo{person}{Matei Zaharia}.} \bibinfo{year}{2020}\natexlab{}.
\newblock \showarticletitle{ColBERT: Efficient and Effective Passage Search via Contextualized Late Interaction over BERT}. In \bibinfo{booktitle}{\emph{Proceedings of the 43rd International ACM SIGIR Conference on Research and Development in Information Retrieval}} (Virtual Event, China) \emph{(\bibinfo{series}{SIGIR '20})}. \bibinfo{publisher}{Association for Computing Machinery}, \bibinfo{address}{New York, NY, USA}, \bibinfo{pages}{39–48}.
\newblock
\showISBNx{9781450380164}
\urldef\tempurl%
\url{https://doi.org/10.1145/3397271.3401075}
\showDOI{\tempurl}


\bibitem[Krishna et~al\mbox{.}(2017)]%
        {krishna2017visual}
\bibfield{author}{\bibinfo{person}{Ranjay Krishna}, \bibinfo{person}{Yuke Zhu}, \bibinfo{person}{Oliver Groth}, \bibinfo{person}{Justin Johnson}, \bibinfo{person}{Kenji Hata}, \bibinfo{person}{Joshua Kravitz}, \bibinfo{person}{Stephanie Chen}, \bibinfo{person}{Yannis Kalantidis}, \bibinfo{person}{Li-Jia Li}, \bibinfo{person}{David~A Shamma}, {et~al\mbox{.}}} \bibinfo{year}{2017}\natexlab{}.
\newblock \showarticletitle{Visual genome: Connecting language and vision using crowdsourced dense image annotations}.
\newblock \bibinfo{journal}{\emph{International journal of computer vision}}  \bibinfo{volume}{123} (\bibinfo{year}{2017}), \bibinfo{pages}{32--73}.
\newblock


\bibitem[Kwiatkowski et~al\mbox{.}(2019)]%
        {kwiatkowski2019natural}
\bibfield{author}{\bibinfo{person}{Tom Kwiatkowski}, \bibinfo{person}{Jennimaria Palomaki}, \bibinfo{person}{Olivia Redfield}, \bibinfo{person}{Michael Collins}, \bibinfo{person}{Ankur Parikh}, \bibinfo{person}{Chris Alberti}, \bibinfo{person}{Danielle Epstein}, \bibinfo{person}{Illia Polosukhin}, \bibinfo{person}{Jacob Devlin}, \bibinfo{person}{Kenton Lee}, {et~al\mbox{.}}} \bibinfo{year}{2019}\natexlab{}.
\newblock \showarticletitle{Natural questions: a benchmark for question answering research}.
\newblock \bibinfo{journal}{\emph{Transactions of the Association for Computational Linguistics}}  \bibinfo{volume}{7} (\bibinfo{year}{2019}), \bibinfo{pages}{453--466}.
\newblock


\bibitem[Lassance et~al\mbox{.}(2021)]%
        {lassance2021composite}
\bibfield{author}{\bibinfo{person}{Carlos Lassance}, \bibinfo{person}{Thibault Formal}, {and} \bibinfo{person}{St\'{e}phane Clinchant}.} \bibinfo{year}{2021}\natexlab{}.
\newblock \showarticletitle{Composite Code Sparse Autoencoders for First Stage Retrieval}. In \bibinfo{booktitle}{\emph{Proceedings of the 44th International ACM SIGIR Conference on Research and Development in Information Retrieval}} (<conf-loc>, <city>Virtual Event</city>, <country>Canada</country>, </conf-loc>) \emph{(\bibinfo{series}{SIGIR '21})}. \bibinfo{publisher}{Association for Computing Machinery}, \bibinfo{address}{New York, NY, USA}, \bibinfo{pages}{2136–2140}.
\newblock
\showISBNx{9781450380379}
\urldef\tempurl%
\url{https://doi.org/10.1145/3404835.3463066}
\showDOI{\tempurl}


\bibitem[Le(2013)]%
        {le2013building}
\bibfield{author}{\bibinfo{person}{Quoc~V Le}.} \bibinfo{year}{2013}\natexlab{}.
\newblock \showarticletitle{Building high-level features using large scale unsupervised learning}. In \bibinfo{booktitle}{\emph{2013 IEEE international conference on acoustics, speech and signal processing}}. IEEE, \bibinfo{pages}{8595--8598}.
\newblock


\bibitem[Li et~al\mbox{.}(2021)]%
        {li2021learning}
\bibfield{author}{\bibinfo{person}{Minghan Li}, \bibinfo{person}{Xialei Liu}, \bibinfo{person}{Joost van~de Weijer}, {and} \bibinfo{person}{Bogdan Raducanu}.} \bibinfo{year}{2021}\natexlab{}.
\newblock \showarticletitle{Learning to Rank for Active Learning: A Listwise Approach}. In \bibinfo{booktitle}{\emph{2020 25th International Conference on Pattern Recognition (ICPR)}}. \bibinfo{pages}{5587--5594}.
\newblock
\urldef\tempurl%
\url{https://doi.org/10.1109/ICPR48806.2021.9412680}
\showDOI{\tempurl}


\bibitem[Lin et~al\mbox{.}(2014)]%
        {lin2014microsoft}
\bibfield{author}{\bibinfo{person}{Tsung-Yi Lin}, \bibinfo{person}{Michael Maire}, \bibinfo{person}{Serge Belongie}, \bibinfo{person}{James Hays}, \bibinfo{person}{Pietro Perona}, \bibinfo{person}{Deva Ramanan}, \bibinfo{person}{Piotr Doll{\'a}r}, {and} \bibinfo{person}{C~Lawrence Zitnick}.} \bibinfo{year}{2014}\natexlab{}.
\newblock \showarticletitle{Microsoft coco: Common objects in context}. In \bibinfo{booktitle}{\emph{Computer Vision--ECCV 2014: 13th European Conference, Zurich, Switzerland, September 6-12, 2014, Proceedings, Part V 13}}. Springer, \bibinfo{pages}{740--755}.
\newblock


\bibitem[Liu et~al\mbox{.}(2019)]%
        {liu2019roberta}
\bibfield{author}{\bibinfo{person}{Yinhan Liu}, \bibinfo{person}{Myle Ott}, \bibinfo{person}{Naman Goyal}, \bibinfo{person}{Jingfei Du}, \bibinfo{person}{Mandar Joshi}, \bibinfo{person}{Danqi Chen}, \bibinfo{person}{Omer Levy}, \bibinfo{person}{Mike Lewis}, \bibinfo{person}{Luke Zettlemoyer}, {and} \bibinfo{person}{Veselin Stoyanov}.} \bibinfo{year}{2019}\natexlab{}.
\newblock \showarticletitle{Roberta: A robustly optimized bert pretraining approach}.
\newblock \bibinfo{journal}{\emph{arXiv preprint arXiv:1907.11692}} (\bibinfo{year}{2019}).
\newblock


\bibitem[Luo et~al\mbox{.}(2022)]%
        {huaishao2022clip4clip}
\bibfield{author}{\bibinfo{person}{Huaishao Luo}, \bibinfo{person}{Lei Ji}, \bibinfo{person}{Ming Zhong}, \bibinfo{person}{Yang Chen}, \bibinfo{person}{Wen Lei}, \bibinfo{person}{Nan Duan}, {and} \bibinfo{person}{Tianrui Li}.} \bibinfo{year}{2022}\natexlab{}.
\newblock \showarticletitle{CLIP4Clip: An empirical study of CLIP for end to end video clip retrieval and captioning}.
\newblock \bibinfo{journal}{\emph{Neurocomputing}}  \bibinfo{volume}{508} (\bibinfo{year}{2022}), \bibinfo{pages}{293--304}.
\newblock
\showISSN{0925-2312}
\urldef\tempurl%
\url{https://doi.org/10.1016/j.neucom.2022.07.028}
\showDOI{\tempurl}


\bibitem[Ma et~al\mbox{.}(2022)]%
        {ma2022ei}
\bibfield{author}{\bibinfo{person}{Haoyu Ma}, \bibinfo{person}{Handong Zhao}, \bibinfo{person}{Zhe Lin}, \bibinfo{person}{Ajinkya Kale}, \bibinfo{person}{Zhangyang Wang}, \bibinfo{person}{Tong Yu}, \bibinfo{person}{Jiuxiang Gu}, \bibinfo{person}{Sunav Choudhary}, {and} \bibinfo{person}{Xiaohui Xie}.} \bibinfo{year}{2022}\natexlab{}.
\newblock \showarticletitle{EI-CLIP: Entity-Aware Interventional Contrastive Learning for E-Commerce Cross-Modal Retrieval}. In \bibinfo{booktitle}{\emph{Proceedings of the IEEE/CVF Conference on Computer Vision and Pattern Recognition (CVPR)}}. \bibinfo{pages}{18051--18061}.
\newblock


\bibitem[Mackenzie et~al\mbox{.}(2020)]%
        {mackenzie2020efficiency}
\bibfield{author}{\bibinfo{person}{Joel Mackenzie}, \bibinfo{person}{Zhuyun Dai}, \bibinfo{person}{Luke Gallagher}, {and} \bibinfo{person}{Jamie Callan}.} \bibinfo{year}{2020}\natexlab{}.
\newblock \showarticletitle{Efficiency Implications of Term Weighting for Passage Retrieval}. In \bibinfo{booktitle}{\emph{Proceedings of the 43rd International ACM SIGIR Conference on Research and Development in Information Retrieval}} (Virtual Event, China) \emph{(\bibinfo{series}{SIGIR '20})}. \bibinfo{publisher}{Association for Computing Machinery}, \bibinfo{address}{New York, NY, USA}, \bibinfo{pages}{1821–1824}.
\newblock
\showISBNx{9781450380164}
\urldef\tempurl%
\url{https://doi.org/10.1145/3397271.3401263}
\showDOI{\tempurl}


\bibitem[Moffat and Zobel(2008)]%
        {moffat2008rank}
\bibfield{author}{\bibinfo{person}{Alistair Moffat} {and} \bibinfo{person}{Justin Zobel}.} \bibinfo{year}{2008}\natexlab{}.
\newblock \showarticletitle{Rank-biased precision for measurement of retrieval effectiveness}.
\newblock \bibinfo{journal}{\emph{ACM Trans. Inf. Syst.}} \bibinfo{volume}{27}, \bibinfo{number}{1}, Article \bibinfo{articleno}{2} (\bibinfo{date}{dec} \bibinfo{year}{2008}), \bibinfo{numpages}{27}~pages.
\newblock
\showISSN{1046-8188}
\urldef\tempurl%
\url{https://doi.org/10.1145/1416950.1416952}
\showDOI{\tempurl}


\bibitem[Mokady et~al\mbox{.}(2021)]%
        {mokady2021clipcap}
\bibfield{author}{\bibinfo{person}{Ron Mokady}, \bibinfo{person}{Amir Hertz}, {and} \bibinfo{person}{Amit~H Bermano}.} \bibinfo{year}{2021}\natexlab{}.
\newblock \showarticletitle{Clipcap: Clip prefix for image captioning}.
\newblock \bibinfo{journal}{\emph{arXiv preprint arXiv:2111.09734}} (\bibinfo{year}{2021}).
\newblock


\bibitem[Nguyen et~al\mbox{.}(2016)]%
        {nguyen2016ms}
\bibfield{author}{\bibinfo{person}{Tri Nguyen}, \bibinfo{person}{Mir Rosenberg}, \bibinfo{person}{Xia Song}, \bibinfo{person}{Jianfeng Gao}, \bibinfo{person}{Saurabh Tiwary}, \bibinfo{person}{Rangan Majumder}, {and} \bibinfo{person}{Li Deng}.} \bibinfo{year}{2016}\natexlab{}.
\newblock \showarticletitle{MS MARCO: A human generated machine reading comprehension dataset}.
\newblock \bibinfo{journal}{\emph{choice}}  \bibinfo{volume}{2640} (\bibinfo{year}{2016}), \bibinfo{pages}{660}.
\newblock


\bibitem[Nie et~al\mbox{.}(2020)]%
        {nie2020dc}
\bibfield{author}{\bibinfo{person}{Ping Nie}, \bibinfo{person}{Yuyu Zhang}, \bibinfo{person}{Xiubo Geng}, \bibinfo{person}{Arun Ramamurthy}, \bibinfo{person}{Le Song}, {and} \bibinfo{person}{Daxin Jiang}.} \bibinfo{year}{2020}\natexlab{}.
\newblock \showarticletitle{DC-BERT: Decoupling Question and Document for Efficient Contextual Encoding}. In \bibinfo{booktitle}{\emph{Proceedings of the 43rd International ACM SIGIR Conference on Research and Development in Information Retrieval}} (Virtual Event, China) \emph{(\bibinfo{series}{SIGIR '20})}. \bibinfo{publisher}{Association for Computing Machinery}, \bibinfo{address}{New York, NY, USA}, \bibinfo{pages}{1829–1832}.
\newblock
\showISBNx{9781450380164}
\urldef\tempurl%
\url{https://doi.org/10.1145/3397271.3401271}
\showDOI{\tempurl}


\bibitem[Nogueira et~al\mbox{.}(2020)]%
        {nogueira2020document}
\bibfield{author}{\bibinfo{person}{Rodrigo Nogueira}, \bibinfo{person}{Zhiying Jiang}, {and} \bibinfo{person}{Jimmy Lin}.} \bibinfo{year}{2020}\natexlab{}.
\newblock \showarticletitle{Document ranking with a pretrained sequence-to-sequence model}.
\newblock \bibinfo{journal}{\emph{arXiv preprint arXiv:2003.06713}} (\bibinfo{year}{2020}).
\newblock


\bibitem[Nogueira et~al\mbox{.}(2019)]%
        {nogueira2019document}
\bibfield{author}{\bibinfo{person}{Rodrigo Nogueira}, \bibinfo{person}{Wei Yang}, \bibinfo{person}{Jimmy Lin}, {and} \bibinfo{person}{Kyunghyun Cho}.} \bibinfo{year}{2019}\natexlab{}.
\newblock \showarticletitle{Document expansion by query prediction}.
\newblock \bibinfo{journal}{\emph{arXiv preprint arXiv:1904.08375}} (\bibinfo{year}{2019}).
\newblock


\bibitem[Oord et~al\mbox{.}(2018)]%
        {oord2018representation}
\bibfield{author}{\bibinfo{person}{Aaron van~den Oord}, \bibinfo{person}{Yazhe Li}, {and} \bibinfo{person}{Oriol Vinyals}.} \bibinfo{year}{2018}\natexlab{}.
\newblock \showarticletitle{Representation learning with contrastive predictive coding}.
\newblock \bibinfo{journal}{\emph{arXiv preprint arXiv:1807.03748}} (\bibinfo{year}{2018}).
\newblock


\bibitem[Plummer et~al\mbox{.}(2015)]%
        {plummer2015flickr30k}
\bibfield{author}{\bibinfo{person}{Bryan~A Plummer}, \bibinfo{person}{Liwei Wang}, \bibinfo{person}{Chris~M Cervantes}, \bibinfo{person}{Juan~C Caicedo}, \bibinfo{person}{Julia Hockenmaier}, {and} \bibinfo{person}{Svetlana Lazebnik}.} \bibinfo{year}{2015}\natexlab{}.
\newblock \showarticletitle{Flickr30k entities: Collecting region-to-phrase correspondences for richer image-to-sentence models}. In \bibinfo{booktitle}{\emph{Proceedings of the IEEE international conference on computer vision}}. \bibinfo{pages}{2641--2649}.
\newblock


\bibitem[Radford et~al\mbox{.}(2021)]%
        {radford2021learning}
\bibfield{author}{\bibinfo{person}{Alec Radford}, \bibinfo{person}{Jong~Wook Kim}, \bibinfo{person}{Chris Hallacy}, \bibinfo{person}{Aditya Ramesh}, \bibinfo{person}{Gabriel Goh}, \bibinfo{person}{Sandhini Agarwal}, \bibinfo{person}{Girish Sastry}, \bibinfo{person}{Amanda Askell}, \bibinfo{person}{Pamela Mishkin}, \bibinfo{person}{Jack Clark}, \bibinfo{person}{Gretchen Krueger}, {and} \bibinfo{person}{Ilya Sutskever}.} \bibinfo{year}{2021}\natexlab{}.
\newblock \showarticletitle{Learning Transferable Visual Models From Natural Language Supervision}. In \bibinfo{booktitle}{\emph{Proceedings of the 38th International Conference on Machine Learning}} \emph{(\bibinfo{series}{Proceedings of Machine Learning Research}, Vol.~\bibinfo{volume}{139})}, \bibfield{editor}{\bibinfo{person}{Marina Meila} {and} \bibinfo{person}{Tong Zhang}} (Eds.). \bibinfo{publisher}{PMLR}, \bibinfo{pages}{8748--8763}.
\newblock
\urldef\tempurl%
\url{https://proceedings.mlr.press/v139/radford21a.html}
\showURL{%
\tempurl}


\bibitem[Radford et~al\mbox{.}(2018)]%
        {radford2018improving}
\bibfield{author}{\bibinfo{person}{Alec Radford}, \bibinfo{person}{Karthik Narasimhan}, \bibinfo{person}{Tim Salimans}, \bibinfo{person}{Ilya Sutskever}, {et~al\mbox{.}}} \bibinfo{year}{2018}\natexlab{}.
\newblock \showarticletitle{Improving language understanding by generative pre-training}.
\newblock  (\bibinfo{year}{2018}).
\newblock


\bibitem[Ray et~al\mbox{.}(2024)]%
        {ray2024cola}
\bibfield{author}{\bibinfo{person}{Arijit Ray}, \bibinfo{person}{Filip Radenovic}, \bibinfo{person}{Abhimanyu Dubey}, \bibinfo{person}{Bryan Plummer}, \bibinfo{person}{Ranjay Krishna}, {and} \bibinfo{person}{Kate Saenko}.} \bibinfo{year}{2024}\natexlab{}.
\newblock \showarticletitle{cola: A Benchmark for Compositional Text-to-image Retrieval}.
\newblock \bibinfo{journal}{\emph{Advances in Neural Information Processing Systems}}  \bibinfo{volume}{36} (\bibinfo{year}{2024}).
\newblock


\bibitem[Reimers and Gurevych(2019)]%
        {reimers2019sentence}
\bibfield{author}{\bibinfo{person}{Nils Reimers} {and} \bibinfo{person}{Iryna Gurevych}.} \bibinfo{year}{2019}\natexlab{}.
\newblock \showarticletitle{Sentence-bert: Sentence embeddings using siamese bert-networks}.
\newblock \bibinfo{journal}{\emph{arXiv preprint arXiv:1908.10084}} (\bibinfo{year}{2019}).
\newblock


\bibitem[Robertson et~al\mbox{.}(1995)]%
        {robertson1995okapi}
\bibfield{author}{\bibinfo{person}{Stephen~E Robertson}, \bibinfo{person}{Steve Walker}, \bibinfo{person}{Susan Jones}, \bibinfo{person}{Micheline~M Hancock-Beaulieu}, \bibinfo{person}{Mike Gatford}, {et~al\mbox{.}}} \bibinfo{year}{1995}\natexlab{}.
\newblock \showarticletitle{Okapi at TREC-3}.
\newblock \bibinfo{journal}{\emph{Nist Special Publication Sp}}  \bibinfo{volume}{109} (\bibinfo{year}{1995}), \bibinfo{pages}{109}.
\newblock


\bibitem[Santhanam et~al\mbox{.}(2021)]%
        {santhanam2021colbertv2}
\bibfield{author}{\bibinfo{person}{Keshav Santhanam}, \bibinfo{person}{Omar Khattab}, \bibinfo{person}{Jon Saad-Falcon}, \bibinfo{person}{Christopher Potts}, {and} \bibinfo{person}{Matei Zaharia}.} \bibinfo{year}{2021}\natexlab{}.
\newblock \showarticletitle{Colbertv2: Effective and efficient retrieval via lightweight late interaction}.
\newblock \bibinfo{journal}{\emph{arXiv preprint arXiv:2112.01488}} (\bibinfo{year}{2021}).
\newblock


\bibitem[Schroff et~al\mbox{.}(2015)]%
        {schroff2015facenet}
\bibfield{author}{\bibinfo{person}{Florian Schroff}, \bibinfo{person}{Dmitry Kalenichenko}, {and} \bibinfo{person}{James Philbin}.} \bibinfo{year}{2015}\natexlab{}.
\newblock \showarticletitle{FaceNet: A Unified Embedding for Face Recognition and Clustering}. In \bibinfo{booktitle}{\emph{Proceedings of the IEEE Conference on Computer Vision and Pattern Recognition (CVPR)}}.
\newblock


\bibitem[Schuhmann et~al\mbox{.}(2021)]%
        {schuhmann2021laion}
\bibfield{author}{\bibinfo{person}{Christoph Schuhmann}, \bibinfo{person}{Richard Vencu}, \bibinfo{person}{Romain Beaumont}, \bibinfo{person}{Robert Kaczmarczyk}, \bibinfo{person}{Clayton Mullis}, \bibinfo{person}{Aarush Katta}, \bibinfo{person}{Theo Coombes}, \bibinfo{person}{Jenia Jitsev}, {and} \bibinfo{person}{Aran Komatsuzaki}.} \bibinfo{year}{2021}\natexlab{}.
\newblock \showarticletitle{Laion-400m: Open dataset of clip-filtered 400 million image-text pairs}.
\newblock \bibinfo{journal}{\emph{arXiv preprint arXiv:2111.02114}} (\bibinfo{year}{2021}).
\newblock


\bibitem[Sohn(2016)]%
        {sohn2016improved}
\bibfield{author}{\bibinfo{person}{Kihyuk Sohn}.} \bibinfo{year}{2016}\natexlab{}.
\newblock \showarticletitle{Improved Deep Metric Learning with Multi-class N-pair Loss Objective}. In \bibinfo{booktitle}{\emph{Advances in Neural Information Processing Systems}}, \bibfield{editor}{\bibinfo{person}{D.~Lee}, \bibinfo{person}{M.~Sugiyama}, \bibinfo{person}{U.~Luxburg}, \bibinfo{person}{I.~Guyon}, {and} \bibinfo{person}{R.~Garnett}} (Eds.), Vol.~\bibinfo{volume}{29}. \bibinfo{publisher}{Curran Associates, Inc.}
\newblock
\urldef\tempurl%
\url{https://proceedings.neurips.cc/paper_files/paper/2016/file/6b180037abbebea991d8b1232f8a8ca9-Paper.pdf}
\showURL{%
\tempurl}


\bibitem[Suarez et~al\mbox{.}(2018)]%
        {suarez2018data}
\bibfield{author}{\bibinfo{person}{Axel Suarez}, \bibinfo{person}{Dyaa Albakour}, \bibinfo{person}{David Corney}, \bibinfo{person}{Miguel Martinez}, {and} \bibinfo{person}{Jos{\'e} Esquivel}.} \bibinfo{year}{2018}\natexlab{}.
\newblock \showarticletitle{A data collection for evaluating the retrieval of related tweets to news articles}. In \bibinfo{booktitle}{\emph{Advances in Information Retrieval: 40th European Conference on IR Research, ECIR 2018, Grenoble, France, March 26-29, 2018, Proceedings 40}}. Springer, \bibinfo{pages}{780--786}.
\newblock


\bibitem[Thakur et~al\mbox{.}(2021)]%
        {thakur2021beir}
\bibfield{author}{\bibinfo{person}{Nandan Thakur}, \bibinfo{person}{Nils Reimers}, \bibinfo{person}{Andreas R{\"u}ckl{\'e}}, \bibinfo{person}{Abhishek Srivastava}, {and} \bibinfo{person}{Iryna Gurevych}.} \bibinfo{year}{2021}\natexlab{}.
\newblock \showarticletitle{Beir: A heterogenous benchmark for zero-shot evaluation of information retrieval models}.
\newblock \bibinfo{journal}{\emph{arXiv preprint arXiv:2104.08663}} (\bibinfo{year}{2021}).
\newblock


\bibitem[Thorne et~al\mbox{.}(2018)]%
        {thorne2018fever}
\bibfield{author}{\bibinfo{person}{James Thorne}, \bibinfo{person}{Andreas Vlachos}, \bibinfo{person}{Christos Christodoulopoulos}, {and} \bibinfo{person}{Arpit Mittal}.} \bibinfo{year}{2018}\natexlab{}.
\newblock \showarticletitle{FEVER: a large-scale dataset for fact extraction and VERification}.
\newblock \bibinfo{journal}{\emph{arXiv preprint arXiv:1803.05355}} (\bibinfo{year}{2018}).
\newblock


\bibitem[Tian et~al\mbox{.}(2020)]%
        {tian2020contrastive}
\bibfield{author}{\bibinfo{person}{Yonglong Tian}, \bibinfo{person}{Dilip Krishnan}, {and} \bibinfo{person}{Phillip Isola}.} \bibinfo{year}{2020}\natexlab{}.
\newblock \showarticletitle{Contrastive multiview coding}. In \bibinfo{booktitle}{\emph{Computer Vision--ECCV 2020: 16th European Conference, Glasgow, UK, August 23--28, 2020, Proceedings, Part XI 16}}. Springer, \bibinfo{pages}{776--794}.
\newblock


\bibitem[Vidit et~al\mbox{.}(2023)]%
        {vidit2023clip}
\bibfield{author}{\bibinfo{person}{Vidit Vidit}, \bibinfo{person}{Martin Engilberge}, {and} \bibinfo{person}{Mathieu Salzmann}.} \bibinfo{year}{2023}\natexlab{}.
\newblock \showarticletitle{CLIP the Gap: A Single Domain Generalization Approach for Object Detection}. In \bibinfo{booktitle}{\emph{Proceedings of the IEEE/CVF Conference on Computer Vision and Pattern Recognition (CVPR)}}. \bibinfo{pages}{3219--3229}.
\newblock


\bibitem[Voorhees et~al\mbox{.}(2021)]%
        {voorhees2021trec}
\bibfield{author}{\bibinfo{person}{Ellen Voorhees}, \bibinfo{person}{Tasmeer Alam}, \bibinfo{person}{Steven Bedrick}, \bibinfo{person}{Dina Demner-Fushman}, \bibinfo{person}{William~R Hersh}, \bibinfo{person}{Kyle Lo}, \bibinfo{person}{Kirk Roberts}, \bibinfo{person}{Ian Soboroff}, {and} \bibinfo{person}{Lucy~Lu Wang}.} \bibinfo{year}{2021}\natexlab{}.
\newblock \showarticletitle{TREC-COVID: constructing a pandemic information retrieval test collection}. In \bibinfo{booktitle}{\emph{ACM SIGIR Forum}}, Vol.~\bibinfo{volume}{54}. ACM New York, NY, USA, \bibinfo{pages}{1--12}.
\newblock


\bibitem[Voorhees et~al\mbox{.}(2003)]%
        {voorhees2003overview}
\bibfield{author}{\bibinfo{person}{Ellen~M Voorhees} {et~al\mbox{.}}} \bibinfo{year}{2003}\natexlab{}.
\newblock \showarticletitle{Overview of the TREC 2003 robust retrieval track.}. In \bibinfo{booktitle}{\emph{Trec}}. \bibinfo{pages}{69--77}.
\newblock


\bibitem[Wadden et~al\mbox{.}(2020)]%
        {wadden2020fact}
\bibfield{author}{\bibinfo{person}{David Wadden}, \bibinfo{person}{Shanchuan Lin}, \bibinfo{person}{Kyle Lo}, \bibinfo{person}{Lucy~Lu Wang}, \bibinfo{person}{Madeleine van Zuylen}, \bibinfo{person}{Arman Cohan}, {and} \bibinfo{person}{Hannaneh Hajishirzi}.} \bibinfo{year}{2020}\natexlab{}.
\newblock \showarticletitle{Fact or fiction: Verifying scientific claims}.
\newblock \bibinfo{journal}{\emph{arXiv preprint arXiv:2004.14974}} (\bibinfo{year}{2020}).
\newblock


\bibitem[Wang et~al\mbox{.}(2022)]%
        {wang2022text}
\bibfield{author}{\bibinfo{person}{Liang Wang}, \bibinfo{person}{Nan Yang}, \bibinfo{person}{Xiaolong Huang}, \bibinfo{person}{Binxing Jiao}, \bibinfo{person}{Linjun Yang}, \bibinfo{person}{Daxin Jiang}, \bibinfo{person}{Rangan Majumder}, {and} \bibinfo{person}{Furu Wei}.} \bibinfo{year}{2022}\natexlab{}.
\newblock \showarticletitle{Text embeddings by weakly-supervised contrastive pre-training}.
\newblock \bibinfo{journal}{\emph{arXiv preprint arXiv:2212.03533}} (\bibinfo{year}{2022}).
\newblock


\bibitem[Yan et~al\mbox{.}(2021)]%
        {yan2021unified}
\bibfield{author}{\bibinfo{person}{Ming Yan}, \bibinfo{person}{Chenliang Li}, \bibinfo{person}{Bin Bi}, \bibinfo{person}{Wei Wang}, {and} \bibinfo{person}{Songfang Huang}.} \bibinfo{year}{2021}\natexlab{}.
\newblock \showarticletitle{A Unified Pretraining Framework for Passage Ranking and Expansion}.
\newblock \bibinfo{journal}{\emph{Proceedings of the AAAI Conference on Artificial Intelligence}} \bibinfo{volume}{35}, \bibinfo{number}{5} (\bibinfo{date}{May} \bibinfo{year}{2021}), \bibinfo{pages}{4555--4563}.
\newblock
\urldef\tempurl%
\url{https://doi.org/10.1609/aaai.v35i5.16584}
\showDOI{\tempurl}


\bibitem[Yu et~al\mbox{.}(2021)]%
        {yu2021improving}
\bibfield{author}{\bibinfo{person}{HongChien Yu}, \bibinfo{person}{Chenyan Xiong}, {and} \bibinfo{person}{Jamie Callan}.} \bibinfo{year}{2021}\natexlab{}.
\newblock \showarticletitle{Improving Query Representations for Dense Retrieval with Pseudo Relevance Feedback}. In \bibinfo{booktitle}{\emph{Proceedings of the 30th ACM International Conference on Information \& Knowledge Management}} (Virtual Event, Queensland, Australia) \emph{(\bibinfo{series}{CIKM '21})}. \bibinfo{publisher}{Association for Computing Machinery}, \bibinfo{address}{New York, NY, USA}, \bibinfo{pages}{3592–3596}.
\newblock
\showISBNx{9781450384469}
\urldef\tempurl%
\url{https://doi.org/10.1145/3459637.3482124}
\showDOI{\tempurl}


\bibitem[Yu et~al\mbox{.}(2022)]%
        {yu2022coca}
\bibfield{author}{\bibinfo{person}{Jiahui Yu}, \bibinfo{person}{Zirui Wang}, \bibinfo{person}{Vijay Vasudevan}, \bibinfo{person}{Legg Yeung}, \bibinfo{person}{Mojtaba Seyedhosseini}, {and} \bibinfo{person}{Yonghui Wu}.} \bibinfo{year}{2022}\natexlab{}.
\newblock \showarticletitle{Coca: Contrastive captioners are image-text foundation models}.
\newblock \bibinfo{journal}{\emph{arXiv preprint arXiv:2205.01917}} (\bibinfo{year}{2022}).
\newblock


\bibitem[Yuan et~al\mbox{.}(2021)]%
        {yuan2021florence}
\bibfield{author}{\bibinfo{person}{Lu Yuan}, \bibinfo{person}{Dongdong Chen}, \bibinfo{person}{Yi-Ling Chen}, \bibinfo{person}{Noel Codella}, \bibinfo{person}{Xiyang Dai}, \bibinfo{person}{Jianfeng Gao}, \bibinfo{person}{Houdong Hu}, \bibinfo{person}{Xuedong Huang}, \bibinfo{person}{Boxin Li}, \bibinfo{person}{Chunyuan Li}, {et~al\mbox{.}}} \bibinfo{year}{2021}\natexlab{}.
\newblock \showarticletitle{Florence: A new foundation model for computer vision}.
\newblock \bibinfo{journal}{\emph{arXiv preprint arXiv:2111.11432}} (\bibinfo{year}{2021}).
\newblock


\bibitem[Zhai et~al\mbox{.}(2023)]%
        {zhai2023sigmoid}
\bibfield{author}{\bibinfo{person}{Xiaohua Zhai}, \bibinfo{person}{Basil Mustafa}, \bibinfo{person}{Alexander Kolesnikov}, {and} \bibinfo{person}{Lucas Beyer}.} \bibinfo{year}{2023}\natexlab{}.
\newblock \showarticletitle{Sigmoid loss for language image pre-training}.
\newblock \bibinfo{journal}{\emph{arXiv preprint arXiv:2303.15343}} (\bibinfo{year}{2023}).
\newblock


\bibitem[Zhao et~al\mbox{.}(2022)]%
        {zhao2022dense}
\bibfield{author}{\bibinfo{person}{Wayne~Xin Zhao}, \bibinfo{person}{Jing Liu}, \bibinfo{person}{Ruiyang Ren}, {and} \bibinfo{person}{Ji-Rong Wen}.} \bibinfo{year}{2022}\natexlab{}.
\newblock \showarticletitle{Dense text retrieval based on pretrained language models: A survey}.
\newblock \bibinfo{journal}{\emph{arXiv preprint arXiv:2211.14876}} (\bibinfo{year}{2022}).
\newblock


\bibitem[Zhou et~al\mbox{.}(2022a)]%
        {zhou2022conditional}
\bibfield{author}{\bibinfo{person}{Kaiyang Zhou}, \bibinfo{person}{Jingkang Yang}, \bibinfo{person}{Chen~Change Loy}, {and} \bibinfo{person}{Ziwei Liu}.} \bibinfo{year}{2022}\natexlab{a}.
\newblock \showarticletitle{Conditional Prompt Learning for Vision-Language Models}. In \bibinfo{booktitle}{\emph{Proceedings of the IEEE/CVF Conference on Computer Vision and Pattern Recognition (CVPR)}}. \bibinfo{pages}{16816--16825}.
\newblock


\bibitem[Zhou et~al\mbox{.}(2022b)]%
        {zhou2022learning}
\bibfield{author}{\bibinfo{person}{Kaiyang Zhou}, \bibinfo{person}{Jingkang Yang}, \bibinfo{person}{Chen~Change Loy}, {and} \bibinfo{person}{Ziwei Liu}.} \bibinfo{year}{2022}\natexlab{b}.
\newblock \showarticletitle{Learning to prompt for vision-language models}.
\newblock \bibinfo{journal}{\emph{International Journal of Computer Vision}} \bibinfo{volume}{130}, \bibinfo{number}{9} (\bibinfo{year}{2022}), \bibinfo{pages}{2337--2348}.
\newblock


\end{thebibliography}

\end{document}